\documentclass[article,twocolumn,nofootinbib]{revtex4-1}
\usepackage[colorlinks=true,linkcolor=blue]{hyperref}%
\usepackage{physics}
\usepackage{graphicx}
\usepackage{ulem}
\usepackage{xcolor}

\begin{document}
		
\title{Global properties of nuclei at finite-temperature within the covariant energy density functional theory}

\author{Ante Ravli\'c}
\email{ravlic@frib.msu.edu}
\affiliation{Facility for Rare Isotope Beams, Michigan State University, East Lansing, Michigan 48824, USA} 
\affiliation{Department of Physics, Faculty of Science, University of Zagreb, Bijeni\v cka c. 32, 10000 Zagreb, Croatia}

\author{Esra Y\"uksel}
\email{e.yuksel@surrey.ac.uk}
\affiliation{Department of Physics, University of Surrey, Guildford, Surrey, GU2 7XH, United Kingdom}

\author{Tamara Nik\v{s}i\'c}
\email{tniksic@phy.hr}
\affiliation{Department of Physics, Faculty of Science, University of Zagreb, Bijeni\v cka c. 32, 10000 Zagreb, Croatia}

\author{Nils Paar}
\email{npaar@phy.hr}
\affiliation{Department of Physics, Faculty of Science, University of Zagreb, Bijeni\v cka c. 32, 10000 Zagreb, Croatia}
	
\date{\today} 
	
\begin{abstract}

In stellar environments nuclei appear at finite temperatures, becoming extremely hot in core-collapse supernovae and neutron star mergers. However, due to theoretical and computational complexity, most model calculations of nuclear properties are performed at zero temperature, while those existing at finite temperatures are limited only to selected regions of the nuclide chart. In this study we perform the global calculation of nuclear properties for even-even $8 \leq Z \leq 104$ nuclei at temperatures in range $0\le T \le 2$ MeV. Calculations are based on the finite temperature relativistic Hartree-Bogoliubov model supplemented by the Bonche-Levit-Vautherin vapor subtraction procedure. We find that near the neutron-drip line the continuum states have significant contribution already at moderate temperature $T\approx 1$ MeV, thus emphasising the necessity of the vapor subtraction procedure. Results include neutron emission lifetimes, quadrupole deformations, neutron skin thickness, proton and neutron pairing gaps, entropy and excitation energy. Up to the temperature $T\approx 1$ MeV nuclear landscape is influenced only moderately by the finite-temperature effects, mainly by reducing the pairing correlations. As the temperature increases further, the effects on nuclear structures become pronounced, reducing both the deformations and the shell effects.  
\end{abstract}

\maketitle

\section{Introduction}

Highly-excited (hot) nuclei can be described as compound nuclei, characterized only by their excitation energy $E^*$ and angular momentum $J$, according to Bohr's hypothesis \cite{bohr1998nuclear}. Such nuclei have no memory regarding their formation, and if in thermodynamical equilibrium, they decay by a slow particle or gamma evaporation, having a short lifetime of the order $10^{-22}$ s. The temperature $T$ defines the statistical decay properties of a compound nucleus. One can find very hot nuclei in extreme stellar environments, such as core-collapse supernovae (CCSNe), where the temperatures exceed $2$ MeV \cite{JANKA200738}. Furthermore, temperatures can be even higher in neutron star mergers, one of the prime candidates for the $r$-process site \cite{Baym_2018}, identified as possible source of a considerable amount of the chemical elements heavier than iron. Therefore, there is a strong motivation for studying the properties of finite nuclei at high temperatures. 

Experimentally, the study of hot nuclei is a very challenging task. Primarily, nuclei at finite temperatures are studied by compound nucleus formation, either by the nuclear fusion reactions \cite{PhysRevC.32.1594}, studying their decay products \cite{PhysRevLett.60.1630}, or by measuring the temperature from relative population of excited states \cite{PhysRevLett.55.177,PhysRevC.35.1695}. Concerning the investigation of decay products, especially important is the study of excited nucleus giant dipole resonance (GDR) decay \cite{PhysRevLett.97.012501,snover1986giant,PhysRevC.84.041304,GAARDHOJE1988261,KICINSKAHABIOR1993225,SURAUD1989294}. Considering a simple complete fusion reaction, the laboratory beam energy can be related to the nuclear masses and excitation energy $E^*$. A compound nucleus is formed, which then decays mostly by the emission of light particles or photons, as dictated by the density of states. Assuming a highly excited nucleus, a simple Fermi gas description of the excitation yields $E^* = aT^2$, where $a$ is the density of states parameter \cite{RevModPhys.9.69}. The decay is described by a level density being proportional to $e^{\sqrt{a E^*}}$. Of course, such a simple model is not realized in practice since the fusion reactions are usually incomplete, complicating the kinematics of the model by introducing the residual nucleus. Furthermore, only at very high temperatures ($T \geq 3$ MeV) is the Fermi relation approximately valid. On the practical side, additional difficulties arise from the necessity of detecting all decayed particles. For temperature to be well defined, the compound nucleus has to be in a metastable state, i.e., thermalized, and decay has to be in equilibrium.

Theoretically, the study of hot nuclei is concerned with either static or dynamic properties. Starting from the mean-field picture of a nucleus, time-evolution of the density operator is achieved by solving the time-dependent finite-temperature Hartree-Fock (TDFT-HF) equations, allowing for the study of the dynamic effects, which are of importance in nuclear reactions such as heavy-ion collisions \cite{PhysRevC.13.1226,SIMENEL201819}. By solving the static FT-HF equations, one obtains the picture of a thermalized compound nucleus at finite temperature. If the temperature is high enough, the main properties of a nucleus at finite temperature can be well described by a simple semi-classical or Thomas-Fermi approximation instead of a quantum treatment \cite{SURAUD1987109,LEVIT1985426}. The finite temperature is usually introduced within either canonical or grand-canonical ensemble through a non-interacting HF density matrix. In the HF basis, diagonal matrix elements correspond to the temperature dependent Fermi-Dirac distribution function. Therefore, the nucleons can scatter above the Fermi level, and temperature smears the Fermi surface. This leads to a non-vanishing number of single-particle states found in the particle continuum. A continuum state is characterized by positive single-particle energy and wavefunctions that do not asymptotically vanish at large distances from the nucleus. Theoretical treatment of continuum states requires special care, either by explicitly constructing the many-body Green's function in the spectral representation \cite{PhysRevC.99.014314,PhysRevC.90.054321} or approximately within the Thomas-Fermi approximation \cite{PhysRevC.84.024311}. Such methods are numerically expensive and impractical for large-scale studies of static nuclear properties. Almost 40 years ago, Bonche, Levit, and Vautherin developed a method that allows for vapor subtraction in a straightforward way \cite{BONCHE1985265,BONCHE1984278}.  It is based on the fact that static FT-HF equations correspond to solving a system consisting of a nucleus surrounded by its external vapor. A prescription is given on how to separate the contribution of the continuum, by subtracting the vapor from the FT-HF solution. Once the vapor is removed, the results of the main observables become independent of the basis size. Later, the subtraction procedure was also justified within the Green's function formalism \cite{PhysRevC.61.064317}. 


The introduction of pairing correlations in the FT-HF can be realized by performing the Bogoliubov transformation of the single-particle operators, yielding the FT-HF Bogoliubov (FT-HFB) equations. Approximately, it can be also realized within the FT-HF Bardeen-Cooper-Schrieffer (FT-HFBCS) theory, where the mean-field and pairing interactions are decoupled \cite{GOODMAN198130}.  It is well known that finite-temperature effects lower the strength of the pairing correlations, which can be expressed as reduction of pairing gaps with increasing temperature, resulting eventually in the pairing collapse at the critical temperature \cite{GOODMAN198130,GOODMAN198145,PhysRevC.34.1942}. Furthermore, temperature influences the single-particle energies, leading to a shape transition from a deformed to a spherical shape. Most nuclei at $T > 3$ MeV, being highly excited, are in a normal state (no pairing correlations) and have a spherical shape \cite{PhysRevC.34.1942,EGIDO198677}. With the advent of the nuclear energy density functional (EDF) theory, systematic calculations of finite-temperature properties across the nuclide chart have become feasible. At zero temperature, a significant amount of work has been done with the non-relativistic EDFs such as Skyrme or Gogny within the HFB theory \cite{erler2012limits,PhysRevLett.102.242501,PhysRevC.81.014303}. On the other hand, results at finite-temperature are somewhat more scarce, restricted to selected nuclei and observables. The investigation was performed on the temperature dependence of neutron skin-thickness \cite{Yuksel2014}, the evolution of paring gaps with temperature \cite{PhysRevC.88.034308, PhysRevC.92.014302}, the influence of temperature on fission barriers \cite{PhysRevC.91.034327,PhysRevC.94.024329}, as well as clustering phenomena \cite{PhysRevC.106.054309}, while the Bonche-Levit-Vautherin (BLV) vapor subtraction was only considered in calculating the neutron emission lifetimes \cite{PhysRevC.90.054316}, and properties of some selected nuclei \cite{LISBOA2010345}.   Recently, a global study of finite-temperature properties has been performed across the nuclide chart within the non-relativistic EDF, however, without treatment of the continuum \cite{YUKSEL2021122238}. The extension to relativistic EDFs is achieved through the relativistic FT-HFB theory (FT-RHFB), or the FT-RHB theory, if the Fock terms are omitted \cite{VRETENAR2005101,NIKSIC2011519}. The starting point in defining the relativistic EDFs is the Lagrangian density in which Dirac particles (nucleons) can be written either as exchanging a set of different mesons (meson-exchange) \cite{PhysRevC.71.024312} or as a sum of bilinear covariants of Dirac fields (point-coupling) \cite{PhysRevC.78.034318}. The ground-state properties at zero-temperature have been thoroughly investigated across the nuclide chart by employing multiple relativistic EDFs in Refs. \cite{AFANASJEV2013680,PhysRevC.89.054320,ZHANG2022101488}. At finite temperature, pairing properties were investigated by assuming spherical nuclei within the FT-RH(F)B theory in Refs. \cite{PhysRevC.88.034308,PhysRevC.96.024304}. Shape transitions were studied at the FT-HF mean-field level \cite{PhysRevC.62.044307} and by using the FT-HBCS theory \cite{PhysRevC.97.054302,PhysRevC.96.054308}, confirming that nuclei collapse to spherical configurations above $T > 3$ MeV. In Ref. \cite{PhysRevC.93.024321} multiple observables were studied by including the BLV vapor subtraction. Only recently, in Ref. \cite{Ravlic2023}, have the finite-temperature drip lines been thoroughly mapped within the relativistic EDF framework using a proper vapor subtraction.

The aim of this work is to conduct a global study of nuclear properties at finite temperatures for
even-even $8 \leq Z \leq 104$ nuclei. Calculations are performed within the framework of the finite temperature relativistic Hartree-Bogoliubov model supplemented by Bonche-Levit-Vautherin vapor subtraction procedure. Axial symmetry is assumed throughout the paper. Calculations are performed with three state-of-the-art relativistic EDFs: meson-exchange DD-ME2 \cite{PhysRevC.71.024312}, point-coupling DD-PC1 \cite{PhysRevC.78.034318} and DD-PCX \cite{PhysRevC.99.034318}. We employ several functionals in order to asses the robustness of our results.

The paper is organized as follows. In Sec. \ref{sec:theory} we present the FT-RHB model supplemented with the vapor subtraction procedure. The importance of the proper continuum treatment is demonstrated in Sec. \ref{sec:continuum}. The large-scale calculation of neutron emission lifetimes is presented in Sec. \ref{sec:neutron_emission}. Global calculations of bulk properties of even-even $8 \leq Z \leq 104$ nuclei at finite-temperature are given in Sec. \ref{sec:bulk}. 
Brief summary and outlook for future studies are presented in Sec. \ref{sec:summary}.
\section{Theoretical formalism}\label{sec:theory}
Nuclei at finite temperature can be treated as open systems that exchange both heat and particles described within the grand-canonical ensemble. Such a system is characterized by its grand-potential $\Omega$. To calculate the thermal properties of nuclei, we employ the relativistic EDF approach \cite{VRETENAR2005101,NIKSIC2011519,NIKSIC20141808}, described by the generalized Bogoliubov-Valatin density $\mathcal{R}$, being a statistical mixture of excited states at finite-temperature \cite{GOODMAN198130,RING1984261}. It assumes a form
\begin{equation}
\mathcal{R} = \begin{pmatrix}
\rho_{k k^\prime} & \kappa_{k k^\prime} \\
- \kappa_{k k^\prime}^*  &1 - \rho_{k k^\prime}^*
\end{pmatrix},
\end{equation}
where $\rho$ is the particle density and $\kappa$ is the pairing tensor. They are defined as thermal averages $\langle \cdot \rangle_T$ of quasiparticle (q.p.) operators \cite{GOODMAN198130,RING1984261}
\begin{equation}
\rho_{k k^\prime} = \langle \beta_{k^\prime}^\dag \beta_k \rangle_T, \quad \kappa_{k k^\prime} = \langle \beta_{k^\prime} \beta_k \rangle_T.
\end{equation}
Set of indices $(k,k^\prime)$ spans a $2M \times 2M$ dimensional space of Bogoliubov quasiparticles (q.p.) ($\beta_k, \beta_k^\dag$), $M$ being the number of q.p. states. To account for a non-vanishing number of particles in the continuum at finite temperature, we have implemented the BLV subtraction procedure (For details about the BLV procedure, we refer the reader to Refs. \cite{BONCHE1985265,BONCHE1984278}). The subtracted grand potential is introduced as
\begin{equation}\label{eq:main-blv}
\bar{\Omega} = \Omega [\mathcal{R} ] - \Omega [\tilde{\mathcal{R}}]+ E_c[\rho_p, \tilde{\rho}_p],
\end{equation}
where $\mathcal{R}$ indicates the generalized density of the Nucleus+Vapor system (Nuc+Vap) and $\tilde{\mathcal{R}}$ corresponds to the vapor-only system (Vap). In order to account for the vapor-nucleus Coulomb interaction, the BLV prescription proposes a form of the Coulomb term $E_c[\rho_p, \tilde{\rho}_p]$ which subtracts the long-range vapor contribution on the nucleus \cite{BONCHE1985265,BONCHE1984278}. Here, $\rho_p, \tilde{\rho}_p$ are the proton particle densities of the Nuc+Vap and Vap systems, respectively. Variation of $\bar{\Omega}$ with $\mathcal{R}$ leads to the FT-RHB equation for the Nuc+Vap system 
\begin{equation}\label{eq:ftrhb_nuc_vap}
\begin{pmatrix}
h - \lambda & \Delta \\
- \Delta^* & -h^*+\lambda
\end{pmatrix} \begin{pmatrix}
U \\
V
\end{pmatrix} = E \begin{pmatrix}
U \\
V
\end{pmatrix},
\end{equation}
where $(U,V)$ is a set of q.p. wavefunctions with energy $E$. On the other hand, by performing variations with $\tilde{\mathcal{R}}$ we get the FT-RHB equation for the Vap system
\begin{equation}\label{eq:ftrhb_vap}
\begin{pmatrix}
\tilde{h} - \lambda & \tilde{\Delta} \\
- \tilde{\Delta}^* & -\tilde{h}^*+\lambda
\end{pmatrix} \begin{pmatrix}
\tilde{U} \\
\tilde{V}
\end{pmatrix} = E \begin{pmatrix}
\tilde{U} \\
\tilde{V}
\end{pmatrix},
\end{equation}
with its corresponding set of wavefunctions $(\tilde{U}, \tilde{V})$ and energies $\tilde{E}$. The chemical potential $\lambda$ is defined to reproduce the total particle number (either neutron or proton)
\begin{equation}\label{eq:chemical_potential}
\int d \boldsymbol{r} (\rho(\boldsymbol{r}) - \tilde{\rho}(\boldsymbol{r})) = N.
\end{equation}
The single-particle Dirac Hamiltonian is labelled by $h, \tilde{h}$, and the pairing field is $\Delta, \tilde{\Delta}$, for the Nuc+Vap and Vap systems, respectively. The Dirac Hamiltonian can be written in terms of the scalar ($S$) and vector ($V$) potentials \cite{Niksic2002}
\begin{equation}
h = - i \boldsymbol{\alpha} \nabla + \beta ( m + S(\boldsymbol{r})) + V(\boldsymbol{r}),
\end{equation}
which depend on the chosen form of the relativistic EDF, $m$ being the bare nucleon mass. For the relativistic meson-exchange (ME) interaction, they are functions of sigma, omega, and rho-meson fields obtained by solving the corresponding Klein-Gordon equations \cite{Niksic2002}. By assuming the point-coupling functionals (PC), where meson propagators are replaced with delta functions, the fields are expanded in terms of scalar ($\rho_s$), vector ($\rho_v$), and isovector ($\rho_{tv}$) densities \cite{PhysRevC.78.034318}. Both interactions include a Coulomb field which satisfies the Poisson equation. The introduction of density-dependent couplings in vertices of relativistic interactions also yields a rearrangement term in vector potential \cite{NIKSIC2011519}. In this work, we employ the density-dependent ME functional DD-ME2 \cite{PhysRevC.71.024312} and two sets of density-dependent point-coupling functionals: DD-PC1 \cite{PhysRevC.78.034318} and DD-PCX \cite{PhysRevC.99.034318}. The difference between the Nuc+Vap Dirac field $h$ and Vap fields $\tilde{h}$ is in the initialization of the scalar and vector fields. While for the Nuc+Vap system we assume an initial Woods-Saxon form of the potentials, the Vap system is only initialized with the Coulomb field. Due to the special treatment of the Coulomb field in Eq. (\ref{eq:main-blv}) within the BLV prescription, the Poisson equation for the Coulomb field $V_c$ of both Nuc+Vap and Vap systems has the form
\begin{equation}\label{eq:coulomb}
- \nabla^2 V_c = e (\rho_p(\boldsymbol{r}) - \tilde{\rho}_p(\boldsymbol{r})).
\end{equation}
Such a term results in a coupling between the Nuc+Vap and Vap FT-RHB equations.
The pairing field is calculated as 
\begin{equation}
\Delta_{l l^\prime} = \frac{1}{2} \sum \limits_{k k^\prime} V^{pp}_{l l^\prime k k^\prime} \kappa_{k k^\prime}, \quad \tilde{\Delta}_{l l^\prime} = \frac{1}{2} \sum \limits_{k k^\prime} V^{pp}_{l l^\prime k k^\prime} \tilde{\kappa}_{k k^\prime},
\end{equation}
for Nuc+Vap and Vap systems, respectively. Here, $V^{pp}$ is the matrix element of the pairing interaction for which we adopt a separable form \cite{PhysRevC.80.024313}
\begin{equation}
V(\boldsymbol{r}_1, \boldsymbol{r}_2, \boldsymbol{r}_1^\prime, \boldsymbol{r}_2^\prime) = - G \delta (\boldsymbol{R} - \boldsymbol{R}^\prime) P(\boldsymbol{r},a) P(\boldsymbol{r}^\prime,a) \frac{1}{2} (1 - P^\sigma ),
\end{equation}
where $\displaystyle \boldsymbol{R} = \frac{1}{2}(\boldsymbol{r}_1 + \boldsymbol{r}_2)$ and $\displaystyle \boldsymbol{r} = \boldsymbol{r}_1 - \boldsymbol{r}_2$ denote the centre-of-mass and relative coordinate, respectively, while 
$\displaystyle P(\boldsymbol{r},a)=\frac{1}{(4\pi a^2)^{3/2}}e^{-\boldsymbol{r}^2/4a^2}$ is the Gaussian form-factor. Parameters $G$ and $a$ describe strength and range of the pairing interaction. For DD-ME2 and DD-PC1 parameterizations we use values defined in Ref. \cite{PhysRevC.80.024313}, while for the DD-PCX interaction, parameters from Ref. \cite{PhysRevC.99.034318} are used.

\begin{figure*}
\centering
\includegraphics[width=\linewidth]{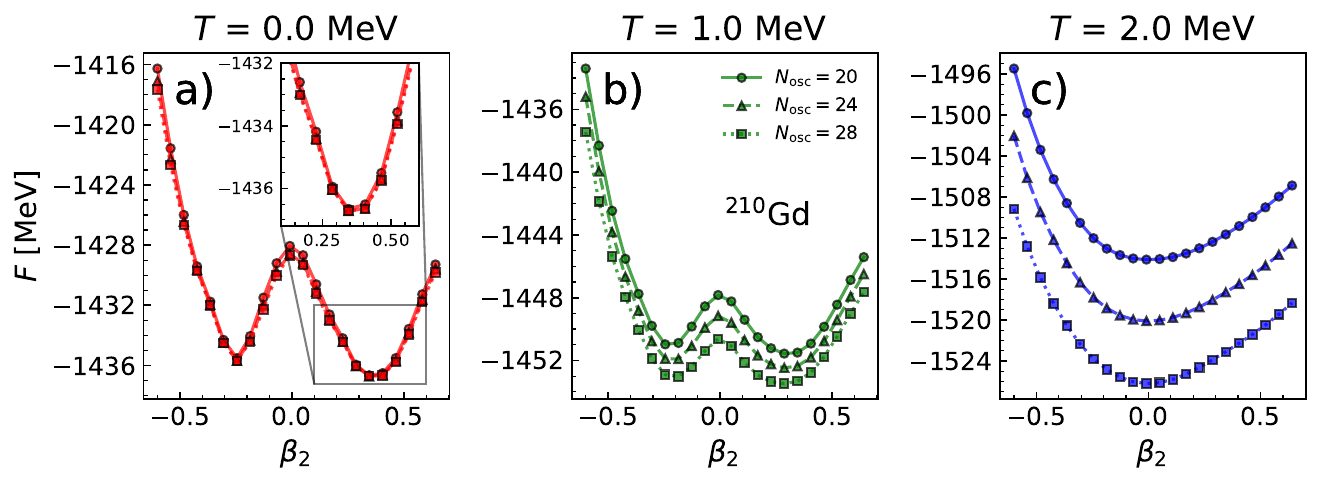}
\caption{(a)-(c) Potential energy curves $(F,\beta_2)$ of ${}^{210}$Gd, calculated for $T = 0,1$ and 2 MeV. Results are shown for different numbers of harmonic oscillator shells $N_{osc}$ without the BLV prescription for vapor subtraction procedure. The inset in panel (a) displays enlarged region around the minimum at $T = 0$ MeV. Calculations are performed with the DD-ME2 interaction. }\label{fig:convergence_issue}
\end{figure*}

We assume axially-deformed reflection-symmetric nuclei, for which the time-reversal invariance holds. This means that angular momentum projection on the $z$-axis $\Omega$ together with parity $\pi$ is a good quantum number, with levels $\pm \Omega$ being degenerate. The optimal configuration is obtained by performing the constrained FT-RHB calculations on the quadrupole deformation $\beta_2$, and minimizing the free energy $F$. We use a mesh of 11 equidistant $\beta_2$ points between $\beta_2 = -0.7$ and $\beta_2 = 0.6$. Constrained calculations are performed for the first 20 iterations, after which the constraint is lifted and calculations converge to the corresponding local minima
in the $(F,\beta_2)$ plane. If there are multiple local minima, the one which minimizes the total free energy is selected (global minimum). No proton-neutron mixing is assumed, allowing only for the isovector ($T=1$) component of the $pp$ interaction. The FT-RHB equations (\ref{eq:ftrhb_nuc_vap}) and (\ref{eq:ftrhb_vap}) are solved in a basis of axially-deformed harmonic oscillator expanded in $N_{osc} = 20$ shells for fermion (and boson) states. We have verified that such a basis yields converged binding energies within 1 MeV for neutron-rich nuclei considered in this work. The Coulomb equation (\ref{eq:coulomb}) is solved by direct integration using Green's function approach. Within one self-consistent iteration, the FT-RHB equations are solved twice, for the Nuc+Vap system and Vap system, supplemented with the chemical potential subsidiary condition in Eq. (\ref{eq:chemical_potential}). Global calculations are performed for nuclei in the range $8 \leq Z \leq 104$. Going above $Z = 104$, we notice the existence of super-deformed minima, in agreement with results from Ref. \cite{PhysRevC.89.054320}. Since our model is not suited for the treatment of those states, we perform our calculations up to the limit of $Z = 104$. 

As a finite-size system, the nucleus is also influenced by fluctuations around the thermal average. Although quantal fluctuations are less relevant at finite-temperature, thermal fluctuations can play a significant role in the description of nuclei by removing the sharp transitions in pairing and deformation properties of nuclei \cite{PhysRevLett.61.767, Egido_1993}. If one assumes the Gaussian approximation, the thermal averages of the main observables (e.g., excitation energies, deformation, and pairing) are weighted by corresponding Boltzmann factors over many thermal configurations \cite{PhysRevC.29.1887,PhysRevC.68.034327}. However, performing large-scale calculations by taking into account the thermal fluctuations is currently not feasible. Therefore, we did not take into account thermal fluctuations in our calculations.

\section{Influence of the particle continuum on the weakly-bound nuclei}\label{sec:continuum}

In this section we discuss the convergence issues that originate from to the continuum contribution to the particle density once we introduce the finite temperature. If single-particle states reach the particle continuum, i.e. acquire positive single-particle energy, their energies will not converge with respect to increasing number of basis states. Already at zero-temperature, one faces a similar problem, which occurs if the pairing correlations are improperly treated \cite{PhysRevC.53.2809}. At finite temperature, even without pairing correlations, such a problem is more pronounced since more single-particle states get scattered across the Fermi level. Often, the contribution of particle continuum is said to be significant for nuclei at $T \geq 4$ MeV \cite{BONCHE1984278,SURAUD1987109}. However, as we will demonstrate, nuclei near the neutron drip lines show convergence problems at much lower temperatures. It is imperative to achieve proper convergence of binding energies to determine the drip lines, which are calculated from the neutron separation energies, defined as the difference  between the total binding energies of neighbouring nuclei. 

First, we perform the axially-deformed constrained FT-RHB calculation to determine the potential energy curve (PEC) and obtain the equilibrium state (defined as the minimum of the free energy $F$ at finite temperature). In this example, we choose ${}^{210}$Gd, which is predicted as a drip line nucleus by using the DD-ME2 interaction, and calculate its PEC at $T = 0, 1$, and $2$ MeV. Results are presented in Fig. \ref{fig:convergence_issue}(a)-(c) for several dimensions of oscillator basis: $N_{osc}=20, 24$ and $28$. In Fig. \ref{fig:convergence_issue}(a) calculations are performed at zero-temperature and the PEC converges already using $N_{osc} = 20$. The difference in the binding energy of the predicted minimum at $\beta_2 \sim 0.35$ between 28 and 20 shells is around 50 keV. Small differences in binding energy verify that using 20 shells is enough for our global study at zero temperature. However, in Fig. \ref{fig:convergence_issue}(b) at $T = 1$ MeV, the PEC does not converge since adding four additional oscillator shells results in decrease of energy by more than $1$ MeV. The situation is even more obvious at $T = 2$ MeV in Fig. \ref{fig:convergence_issue}(c), where adding four oscillator shells shifts the position of the minimum by more than $6$ MeV. Therefore, we can conclude that significant problems with convergence appear already at $T=1$ MeV, much lower than $4$ MeV as stated in the literature.

\begin{figure}
\centering
\includegraphics[scale = 0.80]{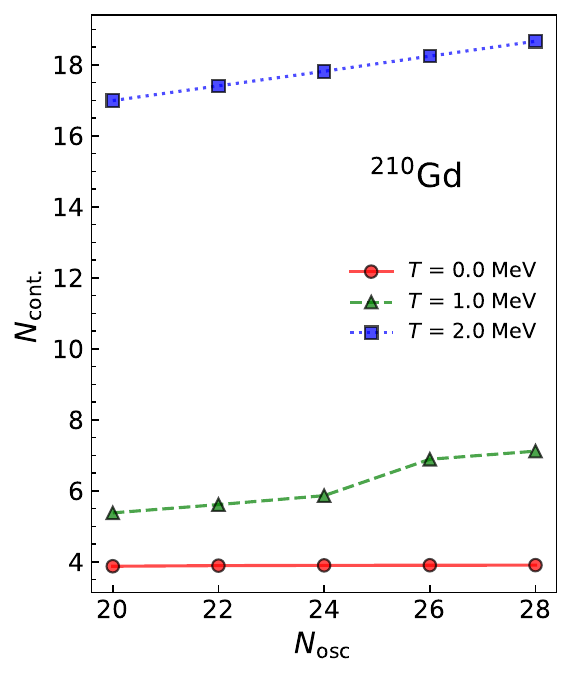}
\caption{Number of neutron continuum states $N_{cont.}$ in ${}^{210}$Gd for different numbers of harmonic oscillator shells $N_{osc}$ at $T = 0,1$ and 2 MeV.}\label{fig:N_cont}
\end{figure}

To illustrate the impact of the continuum states, we calculate the number of states in the continuum $N_{cont.}$. In order to determine the single-particle energy spectrum, we perform the finite-temperature canonical transformation of the q.p. basis. Notice that at finite temperature canonical transformation is just an approximation, since the particle density matrix is not localized \cite{PhysRevC.91.034327,PhysRevC.53.2809}. The number of states in the continuum is defined as the number of single-particle states with positive canonical single-particle energies, $\varepsilon_i > 0$. This amounts to $N_{cont.} = 2\sum \limits_{\varepsilon_i > 0} v_i^2$, where $v_i^2$ denotes the occupation factor in the canonical basis. The results for neutron states of ${}^{210}$Gd, calculated for $\beta_2$ which minimizes the free energy, are displayed in Fig. \ref{fig:N_cont}. Starting from zero temperature, we have found approximately four neutrons in the continuum part of the energy spectrum. However, due to the specific structure of the RHB wavefunctions and continuum coupling, the number of states in the continuum is independent of the basis size and results converge well. Once the temperature is increased, the number of continuum states increases with the basis size. At $T=1$ MeV, the number of neutrons in the continuum increases from approximately $5.4$ for $N_{osc}=20$ shells to approximately $7.1$ for $N_{osc}=28$ shells. Of course, an increasing number of particles in the continuum contributes to the tail region of the particle density. Hence, with increasing dimension of the basis, the density tail grows and observables can display significant dependence on the basis dimension. At $T = 2$ MeV, the number of neutrons in the continuum increases significantly and depends linearly on the number of oscillator shells. The tail region of the density is now even larger and behaves as a nuclear vapor, which inflates with increasing $N_{osc}$. We notice that a similar behavior is found within the BCS theory in the vicinity of neutron drip lines, as demonstrated in Ref. \cite{PhysRevC.53.2809}. Our results clearly show that the proper treatment of continuum is essential in the description of weakly-bound nuclei nearby the drip lines, and special care must be taken at finite temperatures.

\begin{figure}
\centering
\includegraphics[scale = 0.80]{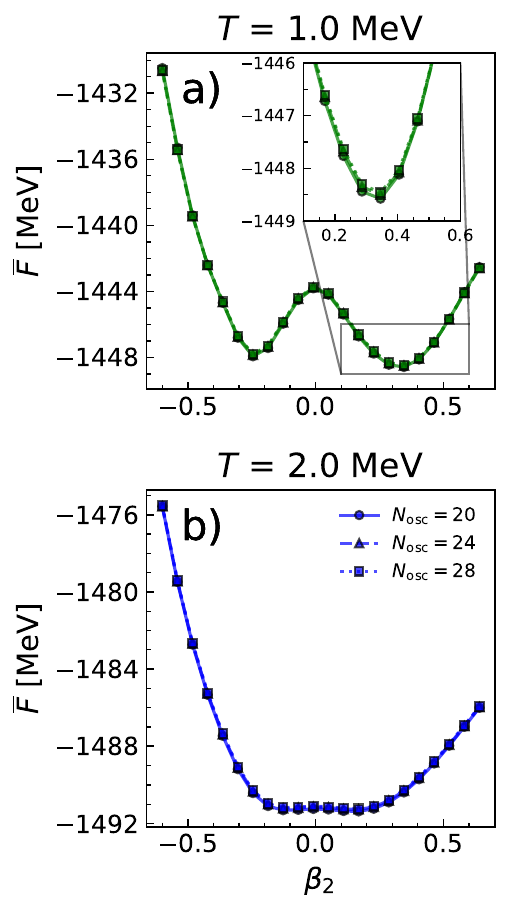}
\caption{(a)-(b) Same as in Fig. \ref{fig:convergence_issue}, shown at temperatures $T = 1$ and 2 MeV, but with subtracted free energy $\bar{F}$ calculated using the BLV prescription. }\label{fig:convergence_issue_no}
\end{figure}

Finally, the converged results can be obtained within the BLV prescription by isolating the continuum states which contribute to nucleon vapor and subtracting them from the calculated observable. In this case, the free energy $F$ is replaced by the subtracted free energy $\bar{F}$, defined as
\begin{equation}
\bar{F} = F_{\text{Nuc+Vap}} - F_{\text{Vap}},
\end{equation}
where $F_{\text{Nuc+Vap(Vap)}}$ and $F_{\text{Vap}}$ are the free energies of the Nuc+Vap and Vap systems, respectively. Results are shown in Fig. \ref{fig:convergence_issue_no}(a)-(b) at $T = 1$ and 2 MeV, and converge very well with respect to the increasing size of the basis. Indeed, since the change in free energy obtained when increasing the basis size from 20 to 28 HO shells is around 100 keV, such results can be used to determine the nuclear drip lines with precision below those occurring due to systematic model uncertainties.

\section{Neutron emission lifetimes}\label{sec:neutron_emission}

Atomic nuclei at finite-temperature are found in highly-excited metastable states that can decay either by particle emission, provided that the excitation energy is above the particle-decay threshold, or by gamma emission \cite{SURAUD1987109}. In particular, as the temperature increases, more and more nuclei gain a finite width for neutron emission. The neutron emission width $\Gamma_n$ can be obtained from the nucleosynthesis formula \cite{BONCHE1984278}
\begin{equation}\label{eq:nucleosynthesys}
\frac{\Gamma_n}{\hbar} = n_{gas} \langle \sigma v \rangle,
\end{equation}
where $\sigma$ is the neutron capture cross-section, $\langle v \rangle$ is the average velocity of particles in the external nucleon gas, and $n_{gas}$ is the neutron vapor density calculated as number of neutrons in the vapor divided by the discretization volume. We approximate the neutron cross-section as $\sigma = \pi R^2$, where the root-mean-square radius $R=\sqrt{\langle r^2\rangle}$ of atomic nucleus is obtained from the FT-RHB calculations. Finally, the neutron emission lifetime can be calculated as $\tau_n = \hbar / \Gamma_n$. The statistical velocity is calculated from the finite-temperature canonical single-particle neutron energies $\varepsilon_n$ assuming the Fermi-Dirac distribution of neutrons $f(\varepsilon_n)$, therefore \cite{PhysRevC.90.054316}
\begin{equation}
\langle v \rangle = \frac{\int \limits_0^\infty f(\varepsilon_n) v (\varepsilon_n) \sqrt{\varepsilon_n} d \varepsilon_n}{\int \limits_0^\infty f(\varepsilon_n) \sqrt{\varepsilon_n} d \varepsilon_n}, \quad v(\varepsilon_n) = \sqrt{\frac{2 \varepsilon_n}{m_n}},
\end{equation}
where $m_n$ is the neutron mass. We obtain the canonical single-particle neutron vapor states by diagonalizing the neutron vapor density $\tilde{\rho}^n$ and transforming the corresponding quasi-particle energies in this basis. This procedure is approximately valid at finite-temperature.

\begin{figure*}
\centering
\includegraphics[width=\linewidth]{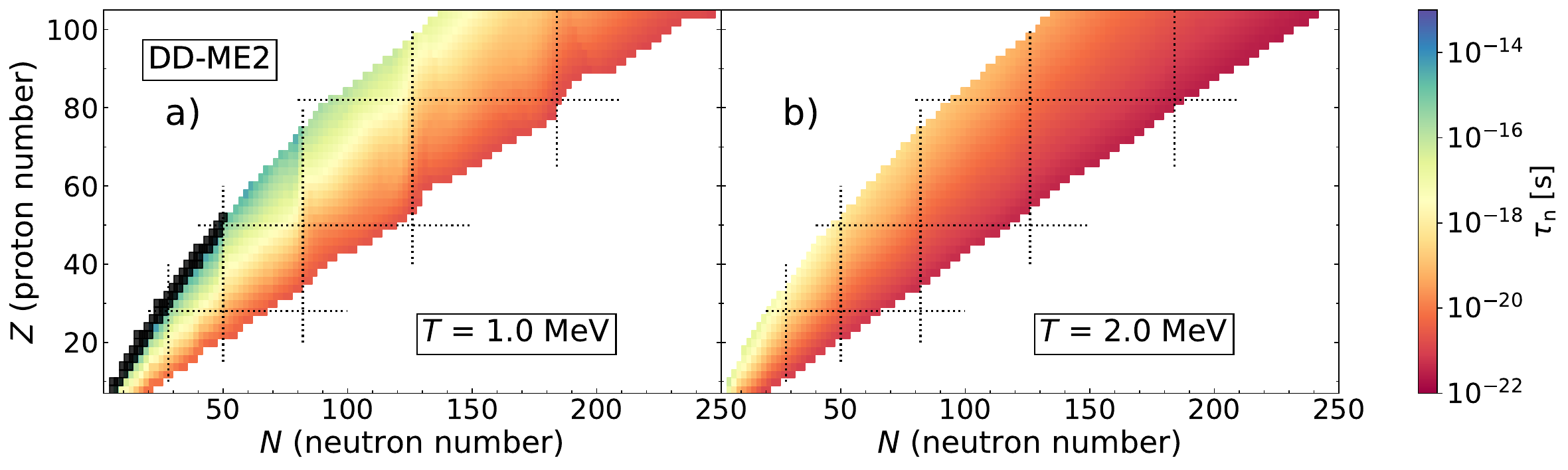}
\caption{Distribution of the neutron emission lifetimes $\tau_n$ for even-even nuclei with proton number in range $8 \leq Z \leq 104$ at temperatures $T = 1$ MeV (a) and $T = 2$ MeV (b). Black dotted lines denote the shell closure numbers, while black squares represent nuclei stable with respect to neutron emission. Calculations are performed with the DD-ME2 interaction.}\label{fig:fig_neutron_emission}
\end{figure*}

We present our calculations for even-even nuclei with $8 \leq Z \leq 104$. Calculations are performed with the DD-ME2 functional at $T = 1$ and 2 MeV and displayed in Fig. \ref{fig:fig_neutron_emission}(a)-(b). We choose relatively high values of temperatures because Eq. (\ref{eq:nucleosynthesys}) is valid for highly-excited nuclei. The density of states for those nuclei is described by a simple Bethe's formula \cite{RevModPhys.9.69}. Indeed, as we will demonstrate later, once both pairing and deformation effects collapse [cf. Sec. \ref{sec:deformation} and \ref{sec:pairing}], the nucleus approximately behaves as a Fermi gas. We calculate the even-even nuclear landscape from the two-proton, up to the two-neutron drip line. The drip lines are defined as \cite{Ravlic2023}

\begin{equation}
S_{2n} = \bar{F}(Z,N) - \bar{F}(Z,N-2), 
\end{equation}
\begin{equation}
S_{2p} = \bar{F}(Z,N)- \bar{F}(Z-2,N),
\end{equation}

where $S_{2n(2p)}$ is the two-neutron(proton) separation energy, and $\bar{F}(Z,N)$ the subtracted free energy of nucleus. This definition is a straightforward generalization of the zero temperature drip line, obtained by substituting the binding energy $E(Z,N)$ with the subtracted free energy $\bar{F}(Z,N) = \bar{E}(Z,N) - T \bar{S}(Z,N)$, where $\bar{S}$ is the subtracted entropy \cite{Ravlic2023}.
In Fig. \ref{fig:fig_neutron_emission}(a) we show the distribution of the neutron emission lifetimes in the nuclide map calculated at $T = 1$ MeV. All even-even nuclei between two-proton and two-neutron drip lines are included. It is interesting to notice that nuclei on the proton-rich side of the nuclide map also acquire a finite width for neutron emission at $T = 1$ MeV. Only a handful of proton-rich nuclei with $Z \leq 52$ are stable against neutron emission (shown as black squares). As the neutron number is increased, the neutron emission widths also increase, reducing the lifetimes by many orders of magnitude. This result is easy to explain in terms of more neutrons being scattered into the vapor states, thus increasing the neutron vapor density. At $T = 1$ MeV, pairing effects collapse, while a significant number of nuclei still display deformation properties. The shell effects are still present thus increasing the stability against neutron emission compared to neighboring nuclei. The two-neutron drip line nuclei have lifetimes of the order $10^{-21}$--${10}^{-22}$ s, comparable to the nuclear thermalization timescale. By increasing the neutron number, $\tau_n$ decreases below the thermalization time, resulting in a non-equilibrated emission of nucleons \cite{BESPROSVANY19891}. To be more precise, the two-neutron drip line at finite temperature should be interpreted as a region that separates the equilibrated neutron emission, from the violent multi-particle emission beyond the finite-temperature drip line. At $T = 2$ MeV in Fig. \ref{fig:fig_neutron_emission}(b) the distribution of neutron emission lifetimes across the nuclear chart looks much smoother, because the shell effects vanish at this temperature. The two-neutron drip line is approximately linear function of the neutron number, while the two-proton drip line departs from the simple linear behavior due to Coulomb repulsion. Nuclei at such high temperatures are well described by the hot liquid-drop model, requiring no shell correction terms. Therefore, our results tend to agree with those from Refs. \cite{BESPROSVANY19891,LEVIT1985426} above $T = 2$ MeV. The neutron lifetimes increase almost monotonically with the neutron number. However, in comparison to calculation at $T=1$ MeV, the average neutron emission lifetimes at the neutron drip line are closer to $10^{-22}$s. We notice that at $T = 2$ MeV no stable nuclei exist with respect to neutron emission. 
\begin{figure*}[t!]
 	\centering
 	\includegraphics[width=\linewidth]{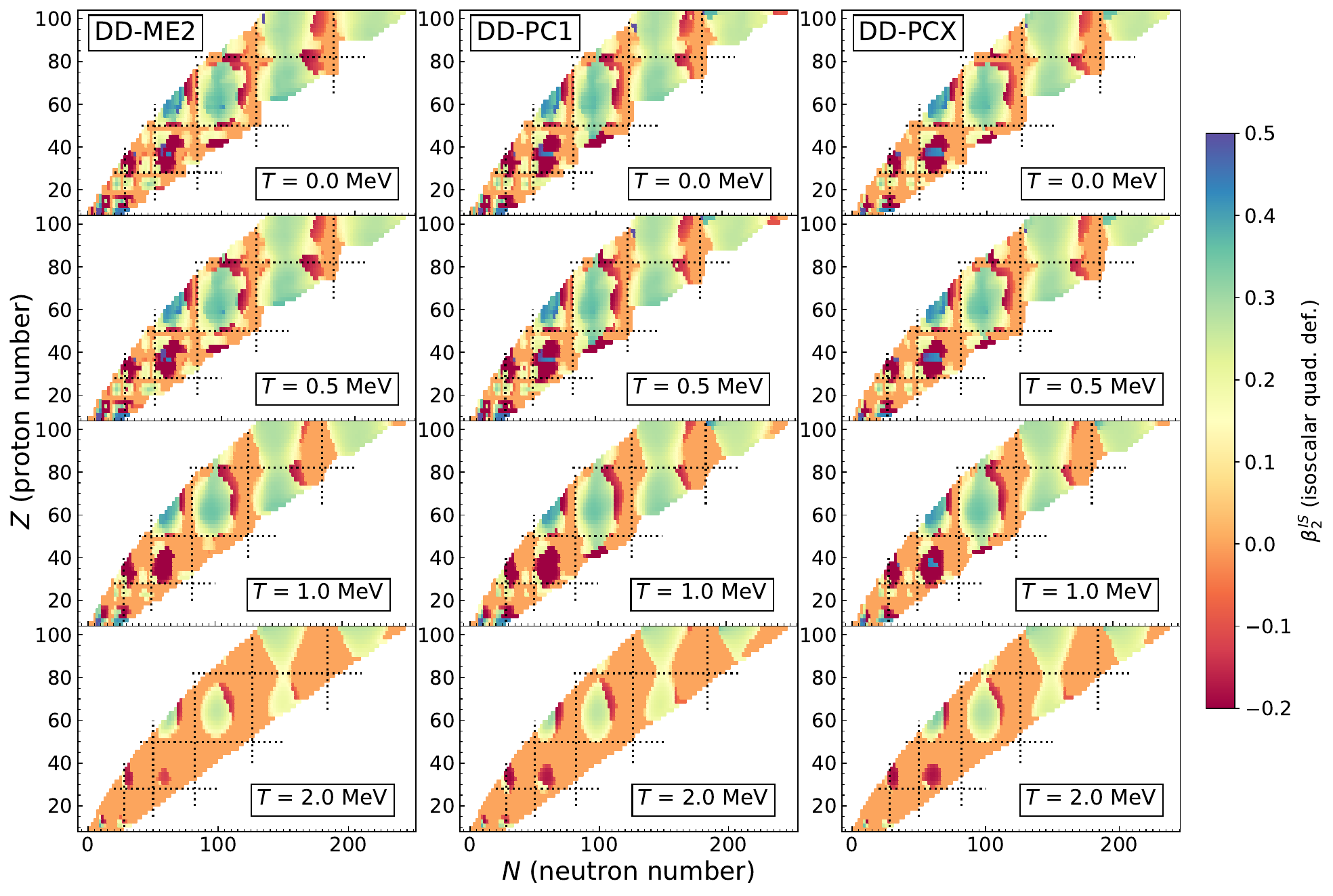}
 	\caption{Distribution of the isoscalar quadrupole deformation $\beta_2^{IS}$ for even-even nuclei with proton number $8\le Z\le 104$ at temperatures $T=0$, $0.5$, $1$ and $2$ MeV. Calculations are performed with DD-ME2 (left panels), DD-PC1 (middle panels) and DD-PCX (right panels) interactions.}\label{fig:deformation}
 \end{figure*}

\section{Selected bulk properties at finite temperature}\label{sec:bulk} 
Within the BLV subtraction procedure, the mean value of an observable $\langle \mathcal{O}[\bar{\rho}] \rangle_T$ at temperature $T$ is a function of the subtracted density ($\bar{\rho}$), defined as the difference between the density of the Nuc+Vap system ($\rho$) and Vap system ($\tilde{\rho}$). For the relativistic EDFs, the baryonic density is equal to the vector density $\bar{\rho}_v$, which satisfies Eq. (\ref{eq:chemical_potential}). In the following, we present the results for the temperature evolution of isoscalar quadrupole deformation, neutron skin-thickness, pairing gap, entropy and excitation energy, for even-even $8 \leq Z \leq 104$ nuclei.

 \subsection{Quadrupole deformation}\label{sec:deformation}
 

 Starting from the proton(neutron) subtracted vector density $\bar{\rho}_v^{p(n)}$ the proton(neutron) quadrupole moment is defined as \cite{PhysRevC.89.054320}
 \begin{equation}
 Q_{20}^{p(n)} = \int d^3 r \bar{\rho}_v^{p(n)}(\boldsymbol{r}) (2z^2 - r_\perp^2),
 \end{equation}
 where $(r_\perp,z)$ are the cylindrical coordinates. It is more customary to express the results in terms of dimensionless variable $\beta_2^{p(n)}$ defined as
\begin{equation}
\beta_2^{p(n)} = \frac{1}{2} \sqrt{\frac{5}{4\pi}} \frac{3}{4\pi} Z(N) R_0^2 Q_{20}^{p(n)},
\end{equation}
where $Z(N)$ denotes the proton(neutron) number and $R_0 = 1.2 A^{1/3}$ fm. The isoscalar quadrupole deformation is defined as $\beta_2^{IS} = \beta_2^p + \beta_2^n$. In Fig. \ref{fig:deformation}, we show the distribution of the isoscalar quadrupole deformation $\beta_2^{IS}$ across the chart of nuclides for three relativistic EDFs employed in this work: DD-ME2, DD-PC1 and DD-PCX. Calculations are performed at temperatures $T=0$, $0.5$, $1$ and $2$ MeV. At $T=0$ MeV we observe spherical shapes in the vicinity of closed shells and deformed for mid-shell nuclei. Temperature effects at $T=0.5$ MeV are too small to alter the shell structure and deformation remains almost unchanged. At $T=1$ MeV, we observe significant increase in number of spherical nuclei and by increasing the temperature further ($T=2$ MeV) most nuclei display spherical shapes except those nuclei with large deformation at $T=0$ MeV.  Apart from small differences mainly for light nuclei, all employed functionals predict similar isoscalar deformations for all temperatures.

\begin{figure}[t!]
    \centering
    \includegraphics[width = 0.9\linewidth]{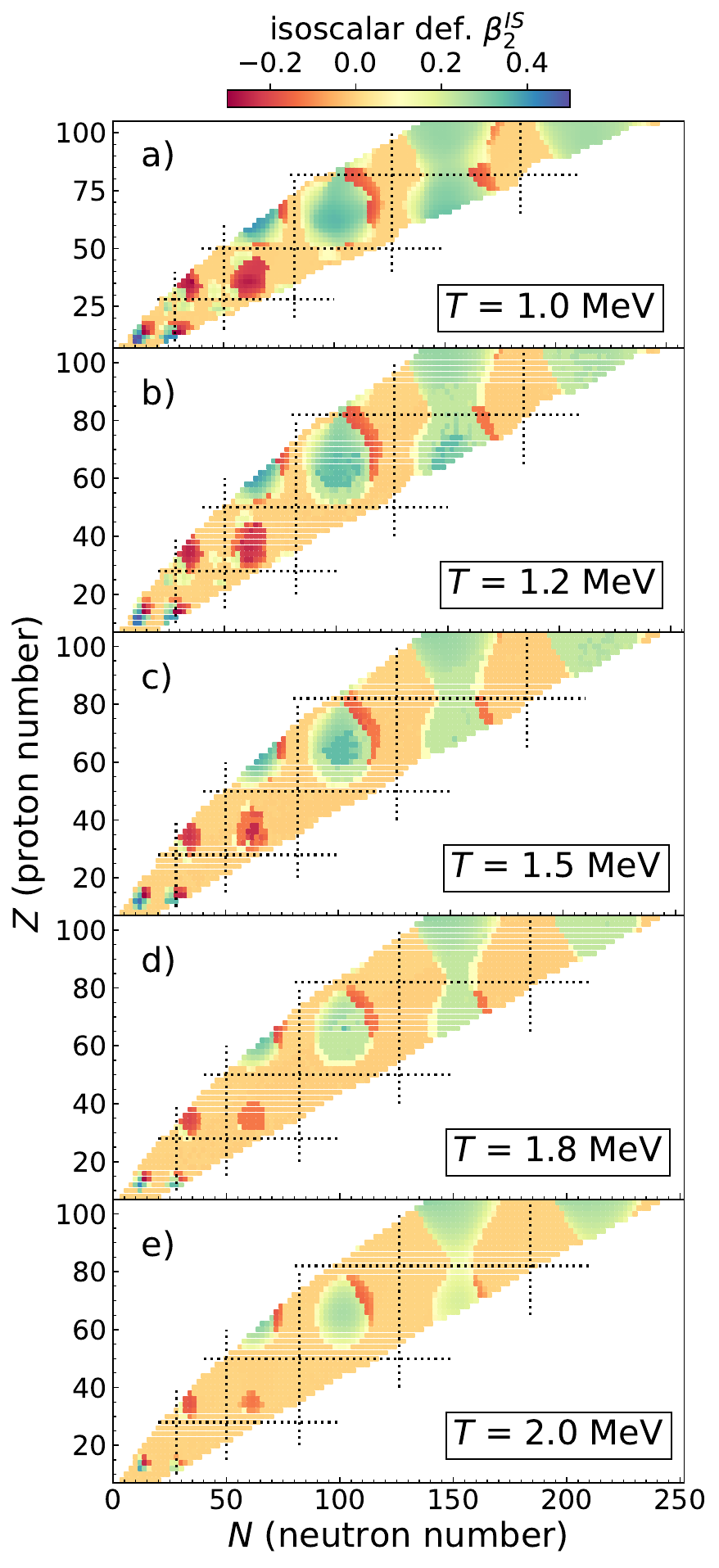}
    \caption{Distribution of the isoscalar quadrupole deformation $\beta_2^{IS}$ for even-even nuclei with proton number $8\le Z\le 104$ on a more refined temperature mesh $T = 1.0, 1.2, 1.5, 1.8$ and $2.0$ MeV. Results are shown for the DD-ME2 interaction.}
    \label{fig:deformation_finer_mesh}
\end{figure}

In order to study sudden change in nuclear shape from $T=1$ MeV to $T=2$ MeV, in Fig. \ref{fig:deformation_finer_mesh}(a)--(e) we show the distribution of $\beta_2^{IS}$ across the chart of nuclides on a more refined temperature mesh:  $T = 1.0$, $1.2$, $1.5$, $1.8$ and $2.0$ MeV. Calculations were performed by using the DD-ME2 interaction, but we notice that both DD-PC1 and DD-PCX interactions follow the similar behavior. The change of nuclear shapes is only moderate up to $T=1$ MeV. However, by further increasing the temperature, it is clearly observed how the islands of axial-deformation gradually reduce in between the shell closure numbers.

 \begin{figure*}[t!]
 	\centering
 	\includegraphics[width=0.45\linewidth]{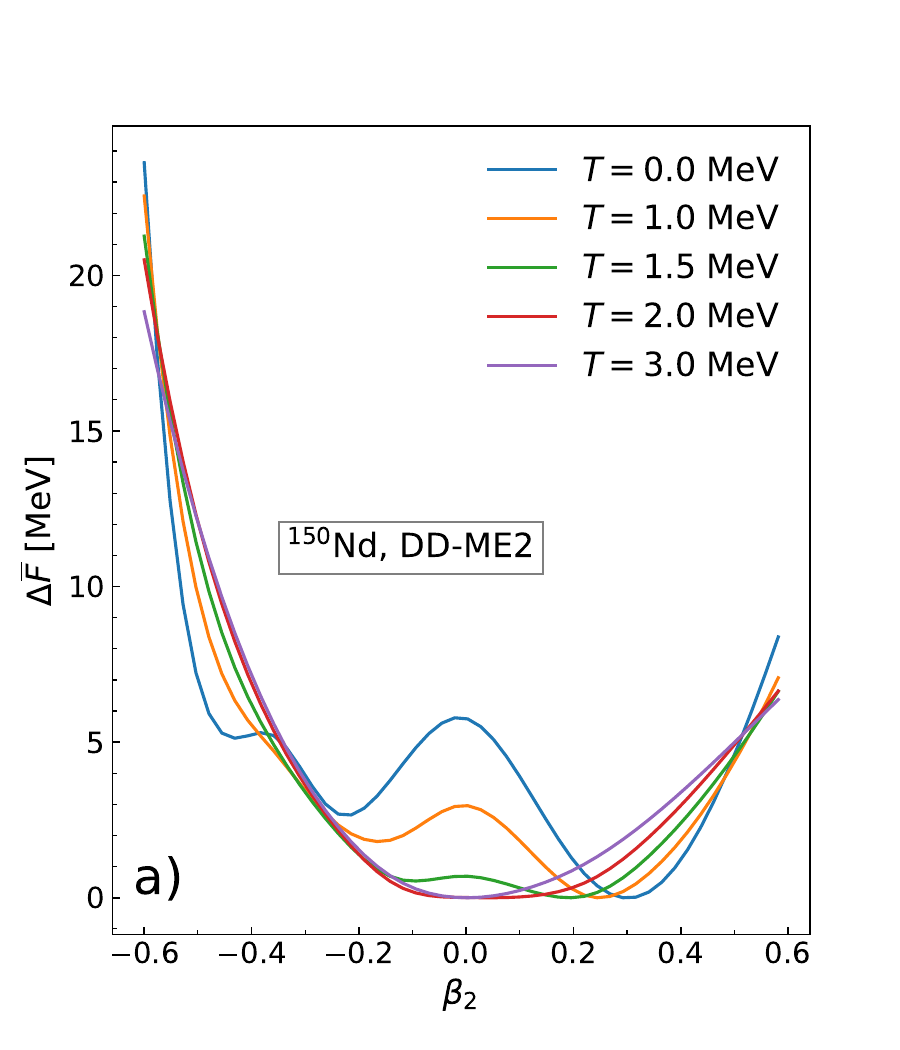}
    \includegraphics[width=0.45\linewidth]{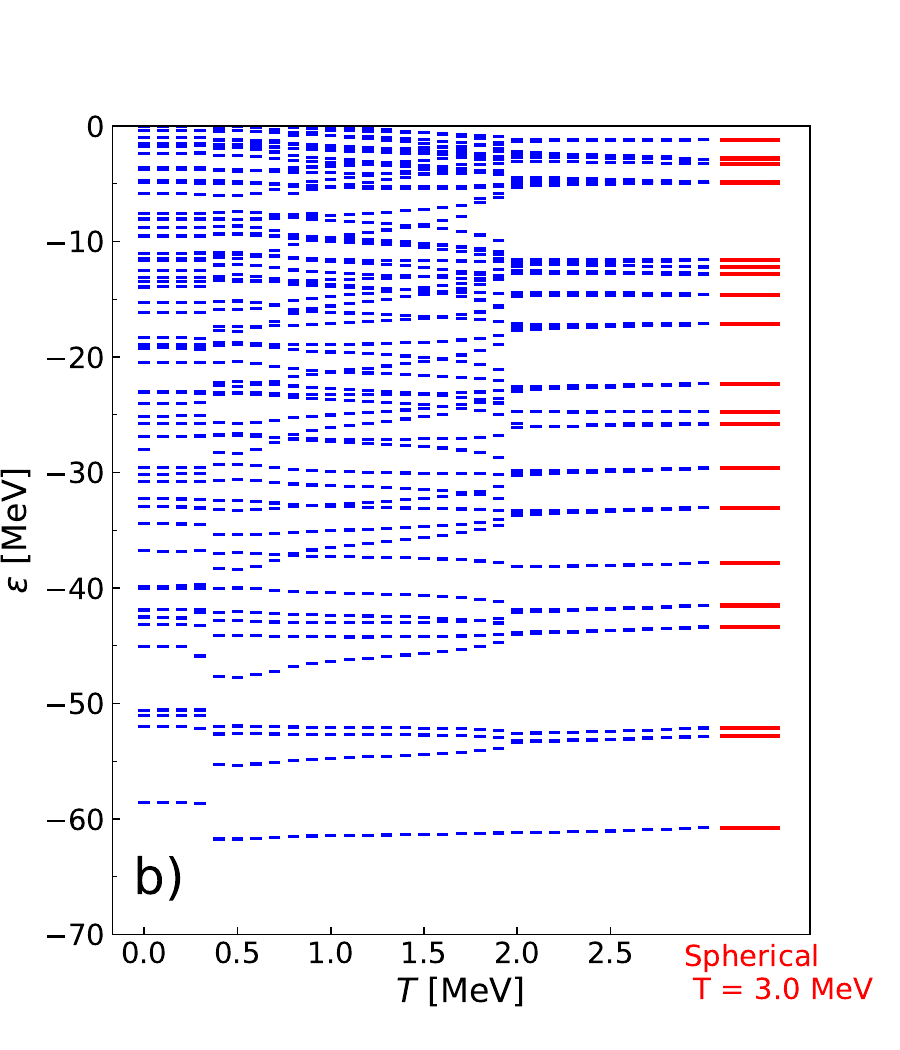}
 	\caption{(a) Potential energy curves of ${}^{150}$Nd isotope for temperatures in range $T = 0$--$3$ MeV calculated with the DD-ME2 interaction. The $\Delta \bar{F}$ represents the relative subtracted free energy with respect to the minimum energy. (b) The single-particle canonical Nuc+Vap states of ${}^{150}$Nd isotope at temperatures $T = 0$--$3$ MeV (blue lines) together with the corresponding spherical states at $T = 3$ MeV (red line). }\label{fig:150nd_pes}
 \end{figure*}

 \begin{figure*}[t!]
 	\centering
 	\includegraphics[width=\linewidth]{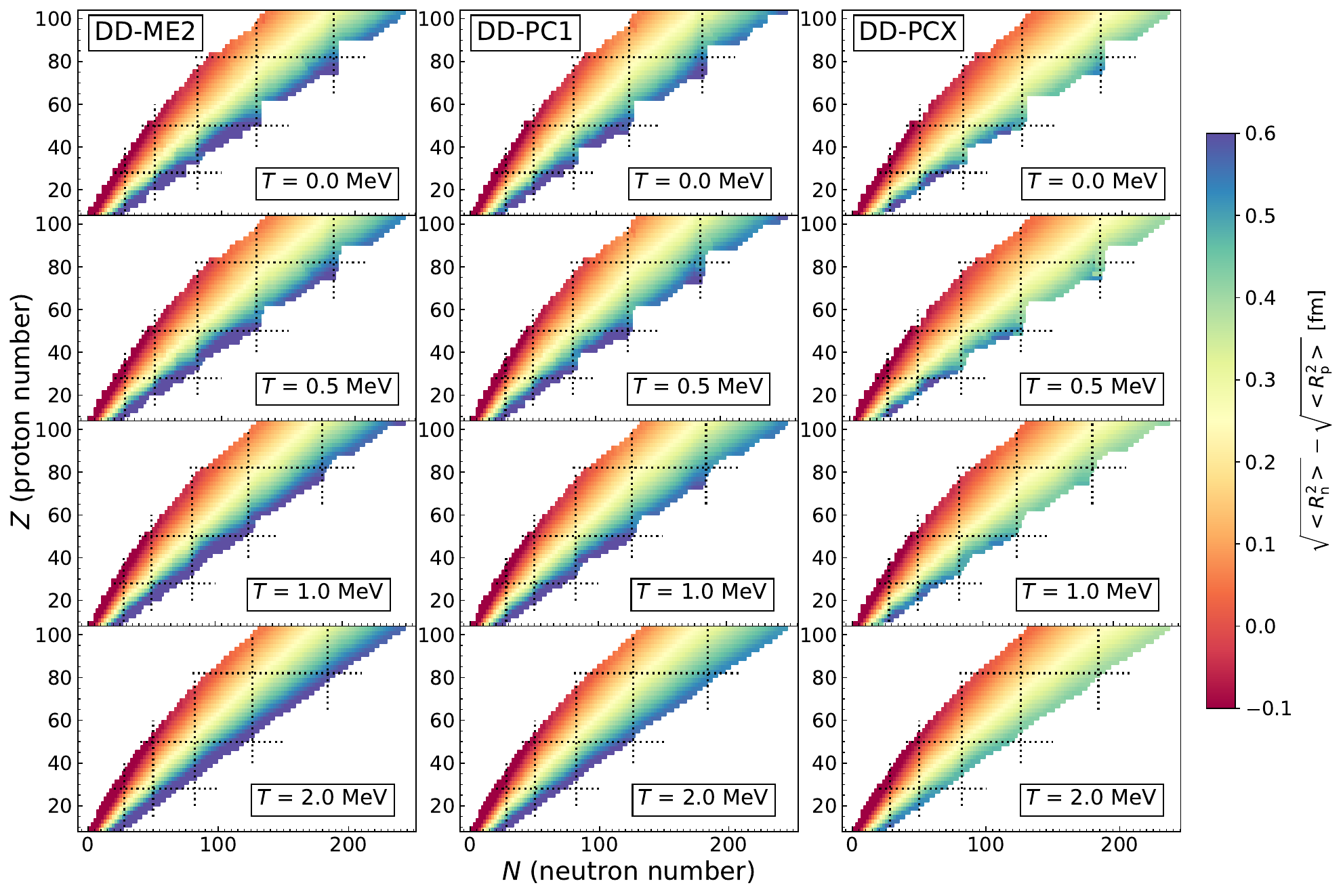}
 	\caption{Same as in Fig. \ref{fig:deformation}, but for the neutron skin thickness, defined as the difference between the neutron and proton root-mean-square radii.}\label{fig:neutron_skin}
 \end{figure*}
 
To investigate the mechanism behind the temperature evolution of quadrupole deformation, we show the PEC for ${}^{150}$Nd in Fig. \ref{fig:150nd_pes}(a), together with deeply bound single-particle canonical states shown in Fig. \ref{fig:150nd_pes}(b). The subtracted free energy in Fig. \ref{fig:150nd_pes}(a) is calculated relative to the global minimum at that temperature and denoted as $\Delta \bar{F}$. Calculations are again performed with the DD-ME2 interaction. At $T=0$ MeV, the PEC for ${}^{150}$Nd displays minima at oblate (located at $\beta_2 = -0.22$) and prolate sides (located at $\beta_2 = 0.29$), with the latter being the global minimum. We note that the spherical configuration ($\beta_2 = 0$) is located approximately 6 MeV above the global minimum. At $T = 1$ MeV, both prolate and oblate minima are found closer to the spherical configuration. The oblate minimum is located at $\beta_2 = -0.17$ and the prolate minimum at $\beta_2 = 0.23$. The excitation energy for the spherical shape decreases by around 3 MeV. By further increasing the temperature, at $T = 2$ MeV, the PEC displays a flat region around $\beta_2=0$, which is a signature of the phase transition. Finally, at $T = 3$ MeV, the minimum is at spherical shape. 
The occurrence of phase shape transitions with increasing temperature can be explained as follows: at finite temperatures, the nucleus gains approximately $k_BT$ additional excitation energy from the environment, and the population of the single-particle levels changes around the Fermi level. At high temperatures, the shell effects disappear, and with the depopulation of the intruder states, which drive the deformation, nuclei become spherical at higher excitation energies \cite{PhysRevLett.85.26,BRACK1974159,PhysRevC.97.054302,LEVIT1984439}.

To study the signature of the phase-transition at the microscopic level, in Fig. \ref{fig:150nd_pes}(b), we also display the single-particle canonical levels of the configuration that minimizes the free energy for temperatures up to $T = 3$ MeV, starting from zero temperature with a step of 0.1 MeV. Canonical single-particle levels up to the continuum threshold ($\varepsilon <0$) are shown in the figure. The canonical single-particle states are characterized by the projection of the total angular momentum on the $z$-axis $\Omega$ and parity $\pi$. Although only an approximation at finite temperatures, the canonical single-particle states suffice to visualise the mechanisms that drive the nucleus to spherical configuration. Starting from low temperatures up to $T=0.5$ MeV, we observe that states corresponding to the same angular momentum $J$ are broken into multiple states represented by the angular momentum projection $\Omega$ and parity $\pi$. The effect of the pairing collapse around $T = 0.5$ MeV clearly leaves a signature on the single-particle levels. As the temperature increases, the energy splitting between these states becomes reduced, finally resulting in restored degeneracy at around $T = 2$ MeV. In the last column in Fig. \ref{fig:150nd_pes}(b), we also show the energy spectrum at $T = 3$ MeV as calculated by imposing the spherical symmetry (red lines). We notice a perfect match between the spherical and axially-deformed calculations at $T = 3$ MeV. Therefore, we anticipate a shape phase-transition in ${}^{150}$Nd at temperatures around 3 MeV.

 \begin{figure*}
 	\centering
 	\includegraphics[width=\linewidth]{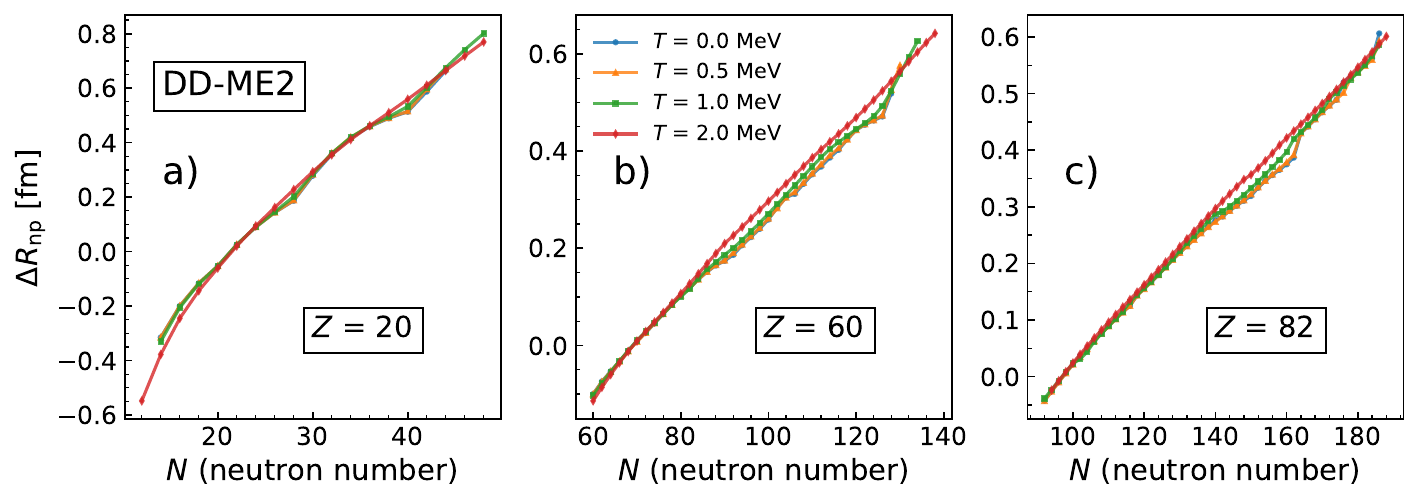}
 	\caption{(a)-(c) The neutron skin thickness $\Delta R_{np}$ as a function of the neutron number for $Z = 20$ (a), $Z = 60$ (b) and $Z = 82$ (c) isotopic chains at temperatures $T = 0$, $0.5$, $1.0$, and $2.0$ MeV. Calculations are performed with the DD-ME2 interaction.}\label{fig:neutron_skin_closeup}
 \end{figure*}


 \subsection{Neutron skin thickness}\label{sec:neutron_skin}
 
 The neutron skin thickness, defined as the difference between the neutron and proton root-mean-square (RMS) radii 
 \begin{equation}
 \Delta R_{np} = \sqrt{\langle R_n^2 \rangle} - \sqrt{\langle R_p^2 \rangle},
 \end{equation}
 provides a direct measure of the isospin asymmetry of the system and is related to the isovector properties of the nuclear matter \cite{PhysRevC.72.064309,PhysRevLett.102.122502,ROCAMAZA201896}.
The neutron(proton) RMS radius is calculated as $\langle R_{n(p)}^2 \rangle = \int \limits_0^\infty d^3 \boldsymbol{r} r^2 \bar{\rho}_v^{n(p)}(\boldsymbol{r})$, where $\bar{\rho}_v^{n(p)}$ denotes the subtracted neutron(proton) vector density.

For nuclei in the vicinity of drip-lines, it is important to properly treat the continuum contribution with increasing temperature. This is especially pronounced for neutron states since there is no Coulomb repulsion to provide a potential barrier for continuum states. Without the vapor subtraction, one would get artificially increasing neutron radii when approaching the drip-line.

The distribution of the neutron skin thickness across the nuclide map, calculated at $T=$ 0, 0.5, 1, and 2 MeV, is shown in Fig. \ref{fig:neutron_skin} for three functionals considered in this work: DD-ME2, DD-PC1, and DD-PCX. First, we observe that the scale is skewed towards positive $\Delta R_{np}$ on the neutron-rich side  compared to the negative $\Delta R_{np}$ on the proton-rich side. This is simply a consequence of the Coulomb repulsion between the protons. Results obtained for the DD-ME2 and DD-PC1 functionals are almost consistent, while the DD-PCX predicts lower $\Delta R_{np}$ for nuclei near the neutron drip-line. Such an outcome is related to different isovector properties among the functionals. It is well established that $\Delta R_{np}$ shows a linear dependence on the symmetry energy ($J$) and its slope ($L$) at saturation density, which is the smallest for DD-PCX. Such a trend is maintained for all temperatures up to $T = 2$ MeV. 

In order to better infer the finite-temperature effects on the neutron skin, in Fig. \ref{fig:neutron_skin_closeup}, we display the temperature dependence of the neutron skin thickness for $Z = 20, 60$ and 82 isotopic chains, calculated with the DD-ME2 interaction. We notice that, for $T=0$, $0.5$ and $1$ MeV, the neutron skin thickness is slightly influenced by the temperature. Only at $T = 2$ MeV, for $Z = 60$ and 82 chains we observe a more pronounced departure from zero-temperature results, especially for larger neutron numbers. This is a consequence of the shape phase-transition which occurs at $T \sim 2$ MeV. Starting from the calcium chain ($Z = 20$) in Fig. \ref{fig:neutron_skin_closeup}(a) for $T\le 1$ MeV, the shell-effects are clearly visible in $\Delta R_{np}$ isotopic dependence, especially around $N = 28$ and $N = 40$, for which our calculations predict pairing collapse. However, as the temperature is increased, the pairing effects are washed-out and isotopic dependence of $\Delta R_{np}$ becomes smoother. At $T = 2$ MeV all calcium isotopes are in a normal state (no pairing correlations) with spherical shape. This leads to linear dependence of $\Delta R_{np}$ on neutron number. The deviations from linear trend for $N<20$ are due to the Coulomb effects. For the neodymium chain (see Fig. \ref{fig:neutron_skin_closeup}(b)), up to $N \approx 82$ the neutron skin thickness is almost temperature independent. For $N > 82$ one can observe temperature effects starting already at $T = 1$ MeV. These nuclei display strongly deformed prolate minima at zero-temperature. As the temperature is increased, their shape changes from prolate deformed to spherical thus causing an almost linear dependence of $\Delta R_{np}$ on neutron number for $T=2$ MeV. We notice that, on the average, the neutron-skin thickness at $T = 2$ MeV is increased compared to lower temperatures, as one would expect. Finite-temperature effects smear the Fermi surface, which leads to the occupation of higher-energy single-particle states, spreading the density tail. Similar trends are also observed for the lead chain in Fig. \ref{fig:neutron_skin_closeup}(c). We notice that in the region between $N = 130$ and $N = 160$, where $\Delta R_{np}$ is not a linear function of $N$, nuclei display prolate shape (see Fig. \ref{fig:deformation}). As the deformation effects are washed-out at $T = 2$ MeV, $\Delta R_{np}$ attains linear dependence on $N$. To conclude our analysis of temperature dependence of neutron-skin thickness, increasing the temperature suppresses the shell effects, thus causing a linear dependence of $\Delta R_{np}$ on neutron number $N$. On average, for neutron-rich nuclei $\Delta R_{np}$ increases with temperature, but only moderately. Our results are also in agreement with the findings in Ref. \cite{PhysRevC.93.024321}, in which proton and neutron radii stay almost constant up to $T = 2$ MeV. Considering the strong correlation between the neutron skin thickness and the slope of the symmetry energy $L$, such a result indicates the stability of the $L$ for temperatures up to $T=2$ MeV, as remarked in Refs. \cite{PhysRevC.85.024310,PhysRevC.95.024314}.

\subsection{Pairing gaps}\label{sec:pairing}

 \begin{figure*}
 	\centering
 	\includegraphics[width=\linewidth]{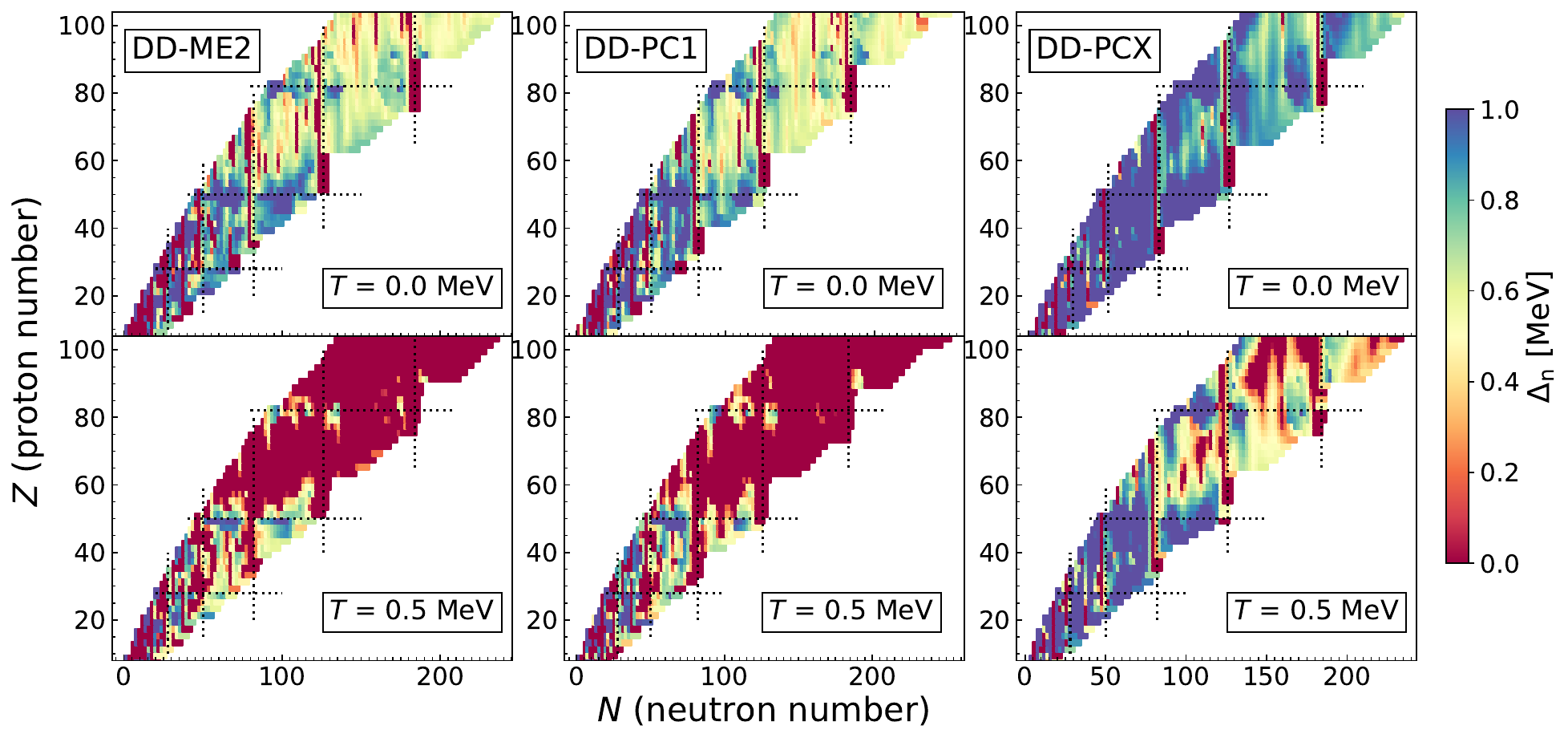}
 	\caption{Distribution of the neutron pairing gap $\Delta_n$ for even-even nuclei with proton number in range $8\le Z\le 104$ at temperatures $T = 0$ and $0.5$ MeV. Calculations are performed with DD-ME2 (left panels), DD-PC1 (middle panels) and DD-PCX (right panels) interactions. }\label{fig:deltan}
 \end{figure*}
 
  \begin{figure*}
 	\centering
 	\includegraphics[width=\linewidth]{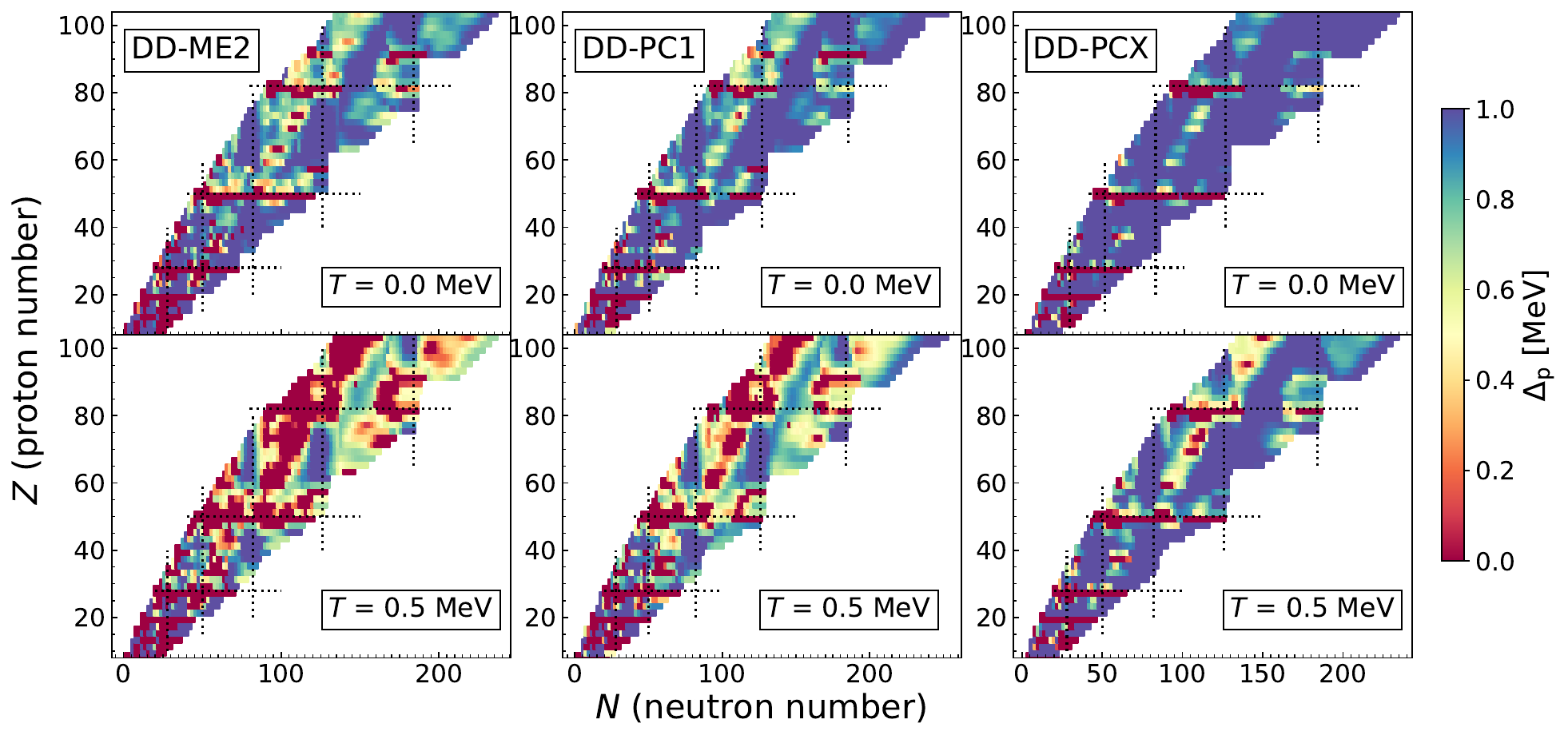}
 	\caption{Same as in Fig. \ref{fig:deltan} but for the proton pairing gaps $\Delta_p$.}\label{fig:deltap}
 \end{figure*}

 \begin{figure*}
 	\centering
 	\includegraphics[width=\linewidth]{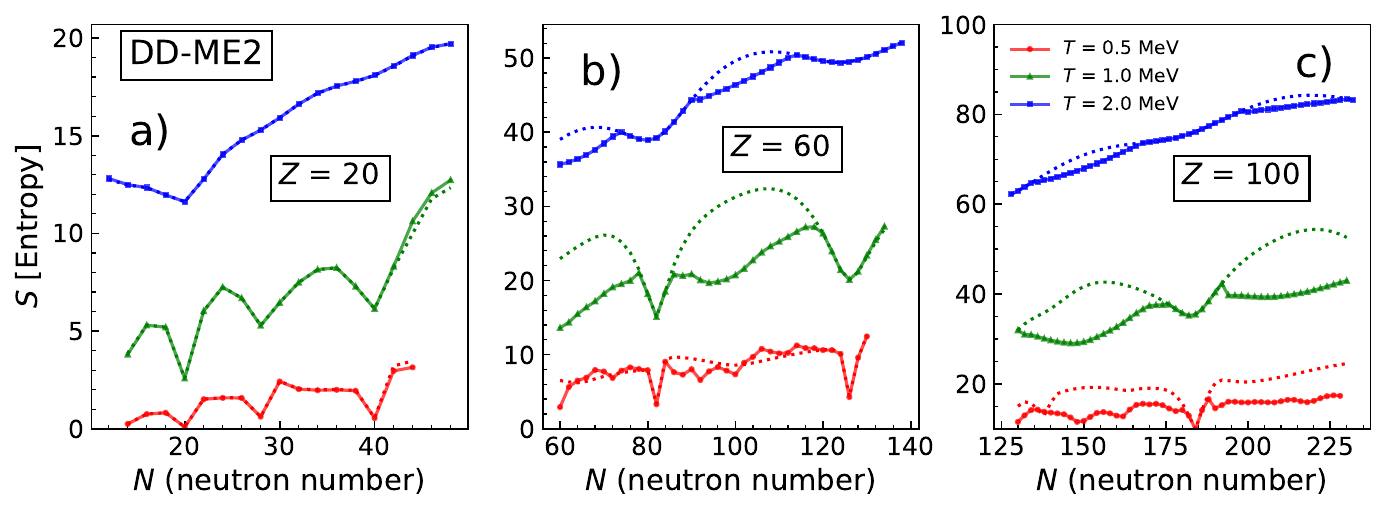}
 	\caption{Entropy $S$ as a function of neutron number for  $Z = 20$(a), $Z = 60$(b) and $Z = 100$(c) isotopic chains. The full line denotes the axially-deformed calculations while the dotted line represents calculations assuming spherical symmetry. Calculations are performed with the DD-ME2 interaction. }\label{fig:entropy_1}
 \end{figure*}
 
 One possible measure of the pairing correlations strength is the neutron(proton) pairing gap $\Delta_{n(p)}$. Although they can be defined in multiple ways \cite{PhysRevC.89.054320}, in this work we employ the definition containing the pairing tensor $\kappa$\footnote{Within the BLV prescription we use the subtracted pairing tensor $\bar{\kappa}$, but since the vapor contribution to pairing is negligible, $\kappa \approx \bar{\kappa}$.}
\begin{equation}
\Delta_{n(p)} = \frac{\sum \limits_{ik} \kappa_{ik} \Delta_{ik}}{\sum \limits_k \kappa_{kk}},
\end{equation}
where $\Delta_{ik}$ is the pairing field. Both the pairing field and the pairing tensor are defined in Refs. \cite{PhysRevC.88.034308,GOODMAN198130}. 

The influence of the finite-temperature on the pairing properties for both relativistic and non-relativistic functionals has been thoroughly investigated in Refs. \cite{PhysRevC.88.034308,GOODMAN198130,PhysRevC.62.044307,Yuksel2014} and therefore we keep our discussions here brief. The main result is that with increasing temperature, one reaches a critical temperature where a phase transition occurs from the superfluid to a normal state. More complex multi-reference calculations as well as the ensemble averaging procedures lead to non-vanishing (although small) pairing gaps \cite{PhysRevC.68.034327,PhysRevC.29.1887}. In this study, we omit thermal averaging because it would be computationally prohibitive for large-scale calculation. For all three functionals employed here (DD-ME2, DD-PC1 and DD-PCX), we use the same separable form of the pairing interaction defined in Refs. \cite{PhysRevC.88.034308,PhysRevC.80.024313}. For both DD-ME2 and DD-PC1 functionals, we use the original values of the pairing interaction parameters $G$ and $a$, while for the DD-PCX functional these parameters have been included in the optimization procedure \cite{PhysRevC.99.034318}. As a result, the pairing strength parameters of the DD-PCX interaction are around 10\% larger compared to the DD-ME2 and DD-PC1.

The distribution of neutron and proton pairing gaps $\Delta_{n(p)}$ across the nuclide map is shown for all three functionals in Figs. \ref{fig:deltan} and \ref{fig:deltap}. Calculations are performed for temperatures $T = 0$ and $0.5$ MeV. We observe that both the neutron and proton pairing gaps vanish in the vicinity of closed shells and increase towards the mid-shell nuclei. On average, the proton pairing gaps are larger in comparison to the neutron pairing gaps. Due to the similar pairing strength, results for DD-ME2 and DD-PC1 functionals are comparable, while the DD-PCX predicts significantly larger pairing gaps. As the temperature increases to $T = 0.5$ MeV, the neutron pairing gaps vanish in a considerable number of nuclei for DD-ME2 and DD-PC1 functionals, while the results calculated with the DD-PCX show a moderate decrease of the neutron pairing gaps with temperature. Further increasing the temperature to $T = 1$ MeV results in neutron pairing collapse for all considered functionals. The proton pairing gaps also decrease when the temperature increases to $T=0.5$ MeV, but not as dramatically as the neutron ones. Again, the DD-PCX shows the least change with temperature due to the higher proton pairing strength. By increasing the temperature to $T = 1$ MeV, only very light nuclei ($Z < 20$) display non-vanishing proton pairing correlations. Therefore, we can conclude that for the vast majority of atomic nuclei, only deformation effects are relevant above $T=1$ MeV, for both proton and neutron states.

 \begin{figure*}
 	\centering
 	\includegraphics[width=\linewidth]{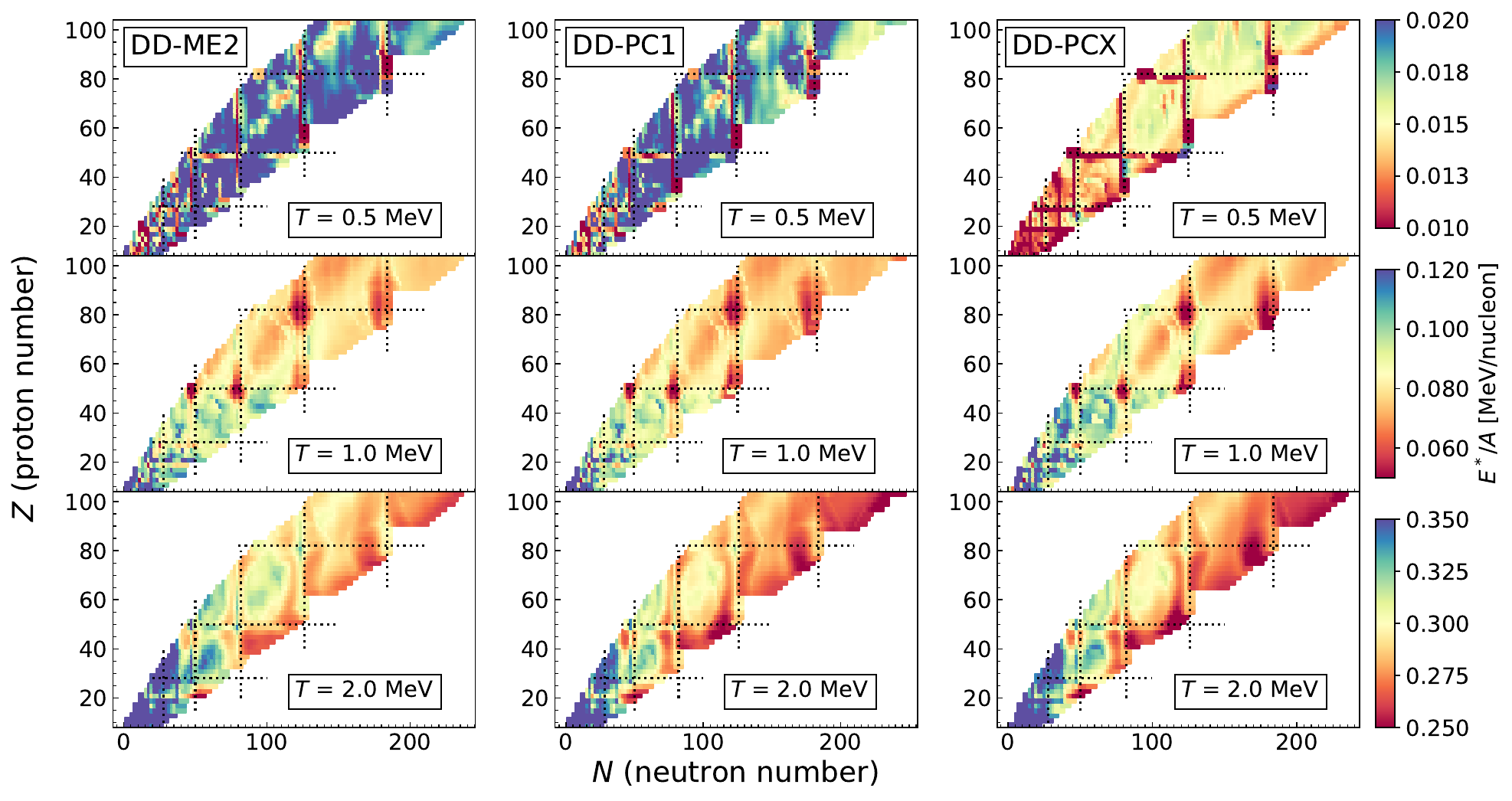}
 	\caption{Distribution of the excitation energy per nucleon $E^*/A$ for even-even nuclei with proton number in range $8\le Z\le 104$ at temperatures $T = 0.5, 1.0$ and $2.0$ MeV. Calculations are performed with DD-ME2 (left panels), DD-PC1 (middle panels) and DD-PCX (right panels) interactions. }\label{fig:excitation_energy}
 \end{figure*}
 
  \begin{figure*}
 	\centering
 	\includegraphics[width=\linewidth]{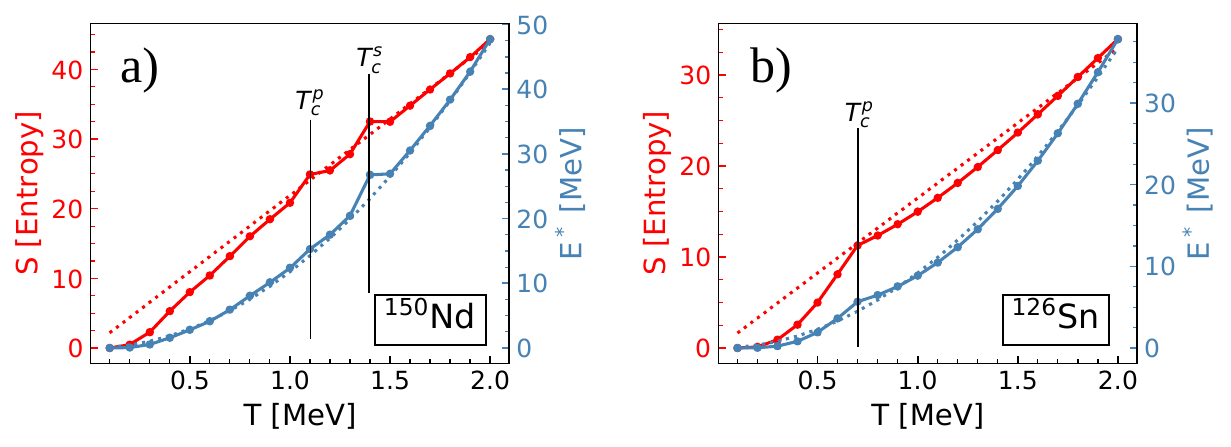}
 	\caption{Entropy $S$ (red line) and excitation energy $E^*$ (blue line) as a function of temperature for ${}^{150}$Nd (a) and ${}^{126}$Sn (b) isotopes. The doted lines denote the fits of the calculated values of entropy and excitation energy to the expression of the Fermi gas model $S=2aT$ and $E^*=aT^2$. Only points with $T\geq 1.5$ MeV are included in the fit.}\label{fig:entropy_2}
 \end{figure*}

\subsection{Entropy and excitation energy}
Unlike the bulk properties discussed in the previous section, entropy is not an observable in the sense that it could be obtained from experiments. Nevertheless, it can provide us with further guidance in interpreting our theoretical calculations. The entropy is a direct measure of occupancy of single-(quasi)particle orbitals and strongly correlates with underlying microscopic structure. It is defined as \cite{GOODMAN198130}
\begin{equation}\label{eq:definition_of_entropy}
    S = - k_B \sum \limits_i \left[f_i \text{ln} f_i + (1-f_i)\text{ln}(1-f_i)\right],
\end{equation}
where $f_i = \left[1 + \text{exp}(\beta E_i)\right]^{-1}$ is the Fermi-Dirac factor for q.p. state with energy $E_i$, and $\beta = 1/k_BT$.

In Figure \ref{fig:entropy_1}(a)-(c), we show the entropy as a function of neutron number for selected isotopic chains $Z=20$, $Z=60$ and $Z=100$, calculated at $T = 0.5, 1$ and $2$ MeV. Full and dotted lines denote calculations assuming axial and spherical symmetry, respectively.
Starting from the calcium isotopic chain in Fig. \ref{fig:entropy_1}(a) at $T = 0.5$ MeV, we observe that the entropy has highly irregular isotopic dependence. Dips observed for $N = 20, 28$ and $40$ correspond to the neutron shell-closures. From the definition of entropy in Eq. (\ref{eq:definition_of_entropy}), it can be inferred that only those levels with semi-occupied shells around the Fermi-level contribute to entropy. For closed shells, all levels are almost fully occupied and only slightly smeared around the Fermi level due to the finite-temperature effects. Therefore, closed shells will be represented as dips when studying the isotopic entropy dependence. As one moves away from closed shells towards the mid-shell, the number of neither completely empty nor completely occupied states increases, resulting in entropy increasing as well. Also shown are the results assuming spherical symmetry (dotted lines). For the calcium isotopic chain, results between spherical and axially-symmetric calculations agree up to the two-neutron drip line. At $T = 1$ MeV, the average entropy of the whole chain increases, however, dips around magic neutron numbers are still pronounced. Again, we note that calculations with assumed spherical symmetry agree well with axially-deformed results up to $N = 42$. In addition to having a spherical shape, pairing effects in calcium isotopes vanish at around $T = 1$ MeV. By further increasing the temperature to $T = 2$ MeV, the entropy curve gets smoother since the temperature is high enough to scatter the nucleons above the closed shells, reducing the dips at magic numbers. This is more pronounced for heavier isotopes, where, due to the neutron excess, nucleons can couple with the continuum more easily. In Fig. \ref{fig:entropy_1}(b) we display the entropy as a function of neutron number in the chain of Nd isotopes. At $T = 0.5$ MeV temperature, two dips in the entropy curve are visible for magic numbers $N = 82$ and $N = 126$. By increasing the temperature to $T = 1$ MeV, it is interesting to notice large differences in entropy between axially-deformed and spherical calculations for mid-shell nuclei that can be explained by the large prolate deformation of Nd isotopes. Since the deformation effects induce degeneracy splitting between different angular momentum projections, there are more states among which the occupation is scattered. This leads to a reduction in entropy compared to the simple spherical geometry, where the entropy is maximum at mid-shell. At $T = 2$ MeV, these differences are still visible, although less pronounced since the deformation splitting of single-particle levels is reduced at higher temperature. For the fermium chain in Fig. \ref{fig:entropy_1}(c), the entropy shows a dip around the $N = 184$ magic number. As the temperature is increased to $T = 1$ MeV, there is a region between $N = 176$--192, where the entropy curve follows the spherical calculation, indicating a widening region of spherical shape around the shell closure number with increasing temperature. At $T = 2$ MeV, this region is even wider, $N = 168$--198, with additional two regions displaying spherical shapes around proton and neutron drip-lines. Therefore, we conclude that deformation effects lead to a reduction of entropy compared to simple spherical calculation, confirming that entropy probes microscopic effects within nuclear structure calculations.

Contrary to entropy, the excitation energy, defined as the difference between the total energy of the atomic nucleus at finite and zero temperature $E^* = E(T) - E(T = 0)$, is accessible in the experiments. The nucleus at finite temperature can be conceptualized as an ensemble average over the excited states, weighted by the Boltzmann factors. Therefore, unlike zero-temperature calculations where the mean-field solution yields the ground state, at finite-temperature we have a mixture of excited states represented by excitation energy $E^*$. In Fig. \ref{fig:excitation_energy}, we show the distribution of the excitation energy per nucleon $E^*/A$ across the nuclide map for three functionals: DD-ME2, DD-PC1 and DD-PCX. Calculations are performed at $T = 0.5, 1$ and $2$ MeV temperatures. Starting from $T = 0.5$ MeV, we observe that the excitation energy varies rapidly across the nuclide map. The nucleon shell-closure numbers can be recognized as dips in excitation energy. By increasing the temperature to $T = 1$ MeV, the doubly-magic nuclei (and nuclei in their vicinity) display smaller values of excitation energy per nucleon in comparison to deformed isotopes. An interesting phenomenon occurs at $T = 2$ MeV where nuclei with shell-closure have larger excitation energies compared to other mid-shell nuclei. Indeed, by examining the results for deformed nuclei at $T = 2$ MeV in Fig. \ref{fig:deformation}, their signature is clearly seen in the lower panel of Fig. \ref{fig:excitation_energy}. Once the energy gap of the closed shells has been surmounted by additional energy, spherical nuclei in the vicinity of shell closure are more easily excited compared to the deformed nuclei. This result is in agreement with the corresponding non-relativistic calculation in Ref. \cite{YUKSEL2021122238}.

 Finally, it is interesting to study the temperature dependence of entropy and excitation energy for some selected nuclei. Calculations are performed for the $^{126}$Sn isotope with a closed proton shell and the mid-shell $^{150}$Nd isotope. Results are shown in Fig. \ref{fig:entropy_2} for temperatures in range $T = 0$--$2$ MeV and calculations are performed by employing the DD-ME2 functional. First, we notice that both entropy and excitation energy increase with temperature, however, the dependence on temperature is not smooth but rather displays visible kinks. For ${}^{150}$Nd shown in Fig. \ref{fig:entropy_2}(a), two such kinks are visible, first at $T_c^p \approx 1.1$ MeV and second at $T_c^s \approx 1.4$ MeV. They correspond to the critical temperature of pairing and shape phase transition, respectively.  On the other hand, in Fig. \ref{fig:entropy_2}(b), ${}^{126}$Sn displays only one kink related to the pairing collapse at $T_c^p \approx 0.7$ MeV. This can be understood by taking into account the spherical shape of the ${}^{126}$Sn isotope for all values of temperature due to the proton shell closure.

Once the pairing and shape effects are washed-out, nucleus behaves approximately as an idealized Fermi gas. Therefore, entropy should be proportional to temperature $S = 2 a T$, while the excitation energy depends quadratically on temperature as $E^* = aT^2$. The constant of proportionality $a$ depends on the density of states as well as the number of nucleons \cite{PhysRev.50.332,bohr1998nuclear}. To compare our results with the Fermi gas model, in Fig. \ref{fig:entropy_2} we include a fit to temperature dependence of $S$ and $E^*$ for the Fermi gas model. Results of the fit for the parameter $a$ using two different definitions are shown in Tab. \ref{tab:a_fit}. Since this model is valid for high temperatures, only temperature values in range $T\ge 1.5$ MeV were included in the fit. The fitted curves are denoted by the dotted lines in  Fig. \ref{fig:entropy_2}. The fitted values of level density parameter $a$, either to entropy or excitation energy, for the same nucleus agree within 10\%. We observe that once $T > T_c$, where $T_c = \max\{T_c^p, T_c^s \}$, both entropy and excitation energy approach the temperature dependence of the Fermi gas model. This results indicate the validity of an independent nucleon picture at high temperatures, when shell effects are diminished. 

\begin{table}[]
    \centering
    \caption{Values of constant $a$ obtained by fitting entropy and excitation energy calculated with the DD-ME2 interaction to temperature dependence of $S$ and $E^*$ for the Fermi gas model, i.e., $S = 2aT$ and $E^* = aT^2$. Fit was performed for two isotopes, ${}^{150}$Nd and ${}^{124}$Sn, leading to results that are consistent within 10\%.}
    \begin{tabular}{c|cc}
    \hline
    \hline
                  & $a({}^{150}\text{Nd})$ [MeV${}^{-1}$] & $a({}^{126}\text{Sn})$ [MeV${}^{-1}$] \\
        \hline
         $S = 2aT$& $10.96 \pm 0.04$ & $8.24 \pm 0.09$  \\
          $E^* = aT^2$  & $11.89 \pm 0.02$     &   $9.23 \pm 0.09$ \\
        \hline
    \end{tabular}
    \label{tab:a_fit}
\end{table}

\section{Summary and outlook}\label{sec:summary}

Finite temperature relativistic Hartree-Bogoliubov model has been supplemented with the vapor subtraction procedure using the BLV prescription. This approach has been employed to study global bulk properties of even-even $8 \leq Z \leq 104$ nuclei.  The importance of the vapor subtraction in weakly-bound nuclei  has been analyzed by studying the particular example of the ${}^{210}$Gd isotope. Without the subtraction procedure, potential energy curve at finite temperature depends sensitively on the size of the basis used to discretize the FT-RHB equation. By subtracting the contribution of the vapor, results become independent of the basis size. 

The bulk properties of nuclei with increasing temperature are mainly influenced by: (i) decrease of pairing gaps, leading to a transition from a superfluid to a normal state, (ii) a shape-phase transition from an axially deformed to a spherical configuration, (iii) reduction of shell gaps. The results of this work can be summarized as follows:

\begin{itemize}
\item The neutron emission lifetimes $\tau_n$ decrease abruptly towards the two-neutron drip line. Furthermore, when increasing the temperature, the neutron emission lifetimes are lower and display smoother behavior across the isotopic chains due to the reduction of shell effects.

\item The isoscalar quadrupole deformations $\beta_2^{IS}$ show visible changes once the temperature is $T \geq 0.5$ MeV. Shape-phase transitions occur initially in nuclei with small deformations and extend towards the mid-shell nuclei as the temperature increases. At temperature around $T = 2$ MeV, most even-even nuclei are predicted to be spherical. 

\item Neutron skin-thickness $\Delta R_{np}$ shows only moderate changes with increasing temperatures. At around $T = 2$ MeV, the isotopic dependence of $\Delta R_{np}$ becomes almost linear for nuclei with neutron excess. On the other hand, proton-rich nuclei show a departure from the linear dependence due to the Coulomb interaction. 

\item The pairing gaps are reduced with increasing temperature. The precise temperature of transition between the superfluid and normal phase depends on the strength of the pairing interaction, but for temperatures $T \geq 1$ MeV, pairing properties vanish for almost all nuclei, except the lightest ones.

\item The isotopic dependence of entropy displays a signature of the underlying microscopic structure. Namely, the entropy decreases towards shell-closure numbers and reaches its peak mid-shell. As the temperature is increased, shell effects are reduced and the isotopic dependence of entropy becomes smooth. In comparison to entropy, excitation energy shows an opposite behavior. Due to the higher density of states in deformed nuclei, their excitation energy is lower as compared to spherical isotopes. Both entropy and excitation energy display kinks near the temperature of the pairing and shape phase transition. At high temperatures, nuclear properties begin to mirror those of an idealized Fermi gas.
\end{itemize}

Once the continuum is properly treated, one could extend the results to nuclei beyond the drip line. Those nuclei would be characterized by a non-equilibrated emission of particles, but their bulk properties at finite-temperature could still be inferred. Furthermore, calculations in this work do not include odd nuclei. Within the mean-field models, odd nuclei are usually treated within the equal-filling approximation (EFA), however, it remains questionable how to extend the EFA to statistical averages within the FT-RHB.  Nevertheless, we leave the consideration of odd nuclei as well as nuclei beyond the two-nucleon drip-line for future work.

\section*{Acknowledgements}

We acknowledge helpful discussions with W. Nazarewicz and S. E. Agbemava. This work is supported by the QuantiXLie Centre of Excellence, a project co financed by the Croatian Government and European Union through the European Regional Development Fund, the Competitiveness and Cohesion Operational Programme (KK.01.1.1.01.0004). This work was supported by the U.S. Department of Energy under Award Number DOE-DE-NA0004074 (NNSA, the Stewardship Science Academic Alliances program) (A.R.). A.R. acknowledges support by the US National Science Foundation under Grant PHY-1927130 (AccelNet-WOU: International Research Network for Nuclear Astrophysics [IReNA]). This work was supported in part through computational resources and services provided by the Institute for Cyber-Enabled Research at Michigan State University. E.Y. acknowledges the support from the Science and Technology Facilities Council (UK) through grant ST/Y000013/1.
	
\bibliographystyle{apsrev4-1}
\bibliography{bibl}

\begin{thebibliography}{73}%
\makeatletter
\providecommand \@ifxundefined [1]{%
 \@ifx{#1\undefined}
}%
\providecommand \@ifnum [1]{%
 \ifnum #1\expandafter \@firstoftwo
 \else \expandafter \@secondoftwo
 \fi
}%
\providecommand \@ifx [1]{%
 \ifx #1\expandafter \@firstoftwo
 \else \expandafter \@secondoftwo
 \fi
}%
\providecommand \natexlab [1]{#1}%
\providecommand \enquote  [1]{``#1''}%
\providecommand \bibnamefont  [1]{#1}%
\providecommand \bibfnamefont [1]{#1}%
\providecommand \citenamefont [1]{#1}%
\providecommand \href@noop [0]{\@secondoftwo}%
\providecommand \href [0]{\begingroup \@sanitize@url \@href}%
\providecommand \@href[1]{\@@startlink{#1}\@@href}%
\providecommand \@@href[1]{\endgroup#1\@@endlink}%
\providecommand \@sanitize@url [0]{\catcode `\\12\catcode `\$12\catcode
  `\&12\catcode `\#12\catcode `\^12\catcode `\_12\catcode `\%12\relax}%
\providecommand \@@startlink[1]{}%
\providecommand \@@endlink[0]{}%
\providecommand \url  [0]{\begingroup\@sanitize@url \@url }%
\providecommand \@url [1]{\endgroup\@href {#1}{\urlprefix }}%
\providecommand \urlprefix  [0]{URL }%
\providecommand \Eprint [0]{\href }%
\providecommand \doibase [0]{http://dx.doi.org/}%
\providecommand \selectlanguage [0]{\@gobble}%
\providecommand \bibinfo  [0]{\@secondoftwo}%
\providecommand \bibfield  [0]{\@secondoftwo}%
\providecommand \translation [1]{[#1]}%
\providecommand \BibitemOpen [0]{}%
\providecommand \bibitemStop [0]{}%
\providecommand \bibitemNoStop [0]{.\EOS\space}%
\providecommand \EOS [0]{\spacefactor3000\relax}%
\providecommand \BibitemShut  [1]{\csname bibitem#1\endcsname}%
\let\auto@bib@innerbib\@empty
\bibitem [{\citenamefont {Bohr}\ and\ \citenamefont
  {Mottelson}(1998)}]{bohr1998nuclear}%
  \BibitemOpen
  \bibfield  {author} {\bibinfo {author} {\bibfnamefont {A.}~\bibnamefont
  {Bohr}}\ and\ \bibinfo {author} {\bibfnamefont {B.}~\bibnamefont
  {Mottelson}},\ }\href {https://books.google.hr/books?id=NNZQDQAAQBAJ} {\emph
  {\bibinfo {title} {Nuclear Structure (In 2 Volumes)}}}\ (\bibinfo
  {publisher} {World Scientific Publishing Company},\ \bibinfo {year}
  {1998})\BibitemShut {NoStop}%
\bibitem [{\citenamefont {Janka}\ \emph {et~al.}(2007)\citenamefont {Janka},
  \citenamefont {Langanke}, \citenamefont {Marek}, \citenamefont
  {Martínez-Pinedo},\ and\ \citenamefont {Müller}}]{JANKA200738}%
  \BibitemOpen
  \bibfield  {author} {\bibinfo {author} {\bibfnamefont {H.-T.}\ \bibnamefont
  {Janka}}, \bibinfo {author} {\bibfnamefont {K.}~\bibnamefont {Langanke}},
  \bibinfo {author} {\bibfnamefont {A.}~\bibnamefont {Marek}}, \bibinfo
  {author} {\bibfnamefont {G.}~\bibnamefont {Martínez-Pinedo}}, \ and\
  \bibinfo {author} {\bibfnamefont {B.}~\bibnamefont {Müller}},\ }\href
  {\doibase https://doi.org/10.1016/j.physrep.2007.02.002} {\bibfield
  {journal} {\bibinfo  {journal} {Physics Reports}\ }\textbf {\bibinfo {volume}
  {442}},\ \bibinfo {pages} {38} (\bibinfo {year} {2007})},\ \bibinfo {note}
  {the Hans Bethe Centennial Volume 1906-2006}\BibitemShut {NoStop}%
\bibitem [{\citenamefont {Baym}\ \emph {et~al.}(2018)\citenamefont {Baym},
  \citenamefont {Hatsuda}, \citenamefont {Kojo}, \citenamefont {Powell},
  \citenamefont {Song},\ and\ \citenamefont {Takatsuka}}]{Baym_2018}%
  \BibitemOpen
  \bibfield  {author} {\bibinfo {author} {\bibfnamefont {G.}~\bibnamefont
  {Baym}}, \bibinfo {author} {\bibfnamefont {T.}~\bibnamefont {Hatsuda}},
  \bibinfo {author} {\bibfnamefont {T.}~\bibnamefont {Kojo}}, \bibinfo {author}
  {\bibfnamefont {P.~D.}\ \bibnamefont {Powell}}, \bibinfo {author}
  {\bibfnamefont {Y.}~\bibnamefont {Song}}, \ and\ \bibinfo {author}
  {\bibfnamefont {T.}~\bibnamefont {Takatsuka}},\ }\href {\doibase
  10.1088/1361-6633/aaae14} {\bibfield  {journal} {\bibinfo  {journal} {Reports
  on Progress in Physics}\ }\textbf {\bibinfo {volume} {81}},\ \bibinfo {pages}
  {056902} (\bibinfo {year} {2018})}\BibitemShut {NoStop}%
\bibitem [{\citenamefont {Jacquet}\ \emph {et~al.}(1985)\citenamefont
  {Jacquet}, \citenamefont {Galin}, \citenamefont {Borderie}, \citenamefont
  {Gardes}, \citenamefont {Guerreau}, \citenamefont {Lefort}, \citenamefont
  {Monnet}, \citenamefont {Rivet}, \citenamefont {Tarrago}, \citenamefont
  {Duek},\ and\ \citenamefont {Alexander}}]{PhysRevC.32.1594}%
  \BibitemOpen
  \bibfield  {author} {\bibinfo {author} {\bibfnamefont {D.}~\bibnamefont
  {Jacquet}}, \bibinfo {author} {\bibfnamefont {J.}~\bibnamefont {Galin}},
  \bibinfo {author} {\bibfnamefont {B.}~\bibnamefont {Borderie}}, \bibinfo
  {author} {\bibfnamefont {D.}~\bibnamefont {Gardes}}, \bibinfo {author}
  {\bibfnamefont {D.}~\bibnamefont {Guerreau}}, \bibinfo {author}
  {\bibfnamefont {M.}~\bibnamefont {Lefort}}, \bibinfo {author} {\bibfnamefont
  {F.}~\bibnamefont {Monnet}}, \bibinfo {author} {\bibfnamefont {M.~F.}\
  \bibnamefont {Rivet}}, \bibinfo {author} {\bibfnamefont {X.}~\bibnamefont
  {Tarrago}}, \bibinfo {author} {\bibfnamefont {E.}~\bibnamefont {Duek}}, \
  and\ \bibinfo {author} {\bibfnamefont {J.~M.}\ \bibnamefont {Alexander}},\
  }\href {\doibase 10.1103/PhysRevC.32.1594} {\bibfield  {journal} {\bibinfo
  {journal} {Phys. Rev. C}\ }\textbf {\bibinfo {volume} {32}},\ \bibinfo
  {pages} {1594} (\bibinfo {year} {1985})}\BibitemShut {NoStop}%
\bibitem [{\citenamefont {Herrmann}\ \emph {et~al.}(1988)\citenamefont
  {Herrmann}, \citenamefont {Bock}, \citenamefont {Emling}, \citenamefont
  {Freifelder}, \citenamefont {Gobbi}, \citenamefont {Grosse}, \citenamefont
  {Hildenbrand}, \citenamefont {Kulessa}, \citenamefont {Matulewicz},
  \citenamefont {Rami}, \citenamefont {Simon}, \citenamefont {Stelzer},
  \citenamefont {Wessels}, \citenamefont {Maurenzig}, \citenamefont {Olmi},
  \citenamefont {Stefanini}, \citenamefont {K\"uhn}, \citenamefont {Metag},
  \citenamefont {Novotny}, \citenamefont {Gnirs}, \citenamefont {Pelte},
  \citenamefont {Braun-Munzinger},\ and\ \citenamefont
  {Moretto}}]{PhysRevLett.60.1630}%
  \BibitemOpen
  \bibfield  {author} {\bibinfo {author} {\bibfnamefont {N.}~\bibnamefont
  {Herrmann}}, \bibinfo {author} {\bibfnamefont {R.}~\bibnamefont {Bock}},
  \bibinfo {author} {\bibfnamefont {H.}~\bibnamefont {Emling}}, \bibinfo
  {author} {\bibfnamefont {R.}~\bibnamefont {Freifelder}}, \bibinfo {author}
  {\bibfnamefont {A.}~\bibnamefont {Gobbi}}, \bibinfo {author} {\bibfnamefont
  {E.}~\bibnamefont {Grosse}}, \bibinfo {author} {\bibfnamefont {K.~D.}\
  \bibnamefont {Hildenbrand}}, \bibinfo {author} {\bibfnamefont
  {R.}~\bibnamefont {Kulessa}}, \bibinfo {author} {\bibfnamefont
  {T.}~\bibnamefont {Matulewicz}}, \bibinfo {author} {\bibfnamefont
  {F.}~\bibnamefont {Rami}}, \bibinfo {author} {\bibfnamefont {R.~S.}\
  \bibnamefont {Simon}}, \bibinfo {author} {\bibfnamefont {H.}~\bibnamefont
  {Stelzer}}, \bibinfo {author} {\bibfnamefont {J.}~\bibnamefont {Wessels}},
  \bibinfo {author} {\bibfnamefont {P.~R.}\ \bibnamefont {Maurenzig}}, \bibinfo
  {author} {\bibfnamefont {A.}~\bibnamefont {Olmi}}, \bibinfo {author}
  {\bibfnamefont {A.~A.}\ \bibnamefont {Stefanini}}, \bibinfo {author}
  {\bibfnamefont {W.}~\bibnamefont {K\"uhn}}, \bibinfo {author} {\bibfnamefont
  {V.}~\bibnamefont {Metag}}, \bibinfo {author} {\bibfnamefont
  {R.}~\bibnamefont {Novotny}}, \bibinfo {author} {\bibfnamefont
  {M.}~\bibnamefont {Gnirs}}, \bibinfo {author} {\bibfnamefont
  {D.}~\bibnamefont {Pelte}}, \bibinfo {author} {\bibfnamefont
  {P.}~\bibnamefont {Braun-Munzinger}}, \ and\ \bibinfo {author} {\bibfnamefont
  {L.~G.}\ \bibnamefont {Moretto}},\ }\href {\doibase
  10.1103/PhysRevLett.60.1630} {\bibfield  {journal} {\bibinfo  {journal}
  {Phys. Rev. Lett.}\ }\textbf {\bibinfo {volume} {60}},\ \bibinfo {pages}
  {1630} (\bibinfo {year} {1988})}\BibitemShut {NoStop}%
\bibitem [{\citenamefont {Pochodzalla}\ \emph {et~al.}(1985)\citenamefont
  {Pochodzalla}, \citenamefont {Friedman}, \citenamefont {Gelbke},
  \citenamefont {Lynch}, \citenamefont {Maier}, \citenamefont {Ardouin},
  \citenamefont {Delagrange}, \citenamefont {Doubre}, \citenamefont
  {Gr\'egoire}, \citenamefont {Kyanowski}, \citenamefont {Mittig},
  \citenamefont {P\'eghaire}, \citenamefont {P\'eter}, \citenamefont
  {Saint-Laurent}, \citenamefont {Viyogi}, \citenamefont {Zwieglinski},
  \citenamefont {Bizard}, \citenamefont {Lef\`ebvres}, \citenamefont {Tamain},\
  and\ \citenamefont {Qu\'ebert}}]{PhysRevLett.55.177}%
  \BibitemOpen
  \bibfield  {author} {\bibinfo {author} {\bibfnamefont {J.}~\bibnamefont
  {Pochodzalla}}, \bibinfo {author} {\bibfnamefont {W.~A.}\ \bibnamefont
  {Friedman}}, \bibinfo {author} {\bibfnamefont {C.~K.}\ \bibnamefont
  {Gelbke}}, \bibinfo {author} {\bibfnamefont {W.~G.}\ \bibnamefont {Lynch}},
  \bibinfo {author} {\bibfnamefont {M.}~\bibnamefont {Maier}}, \bibinfo
  {author} {\bibfnamefont {D.}~\bibnamefont {Ardouin}}, \bibinfo {author}
  {\bibfnamefont {H.}~\bibnamefont {Delagrange}}, \bibinfo {author}
  {\bibfnamefont {H.}~\bibnamefont {Doubre}}, \bibinfo {author} {\bibfnamefont
  {C.}~\bibnamefont {Gr\'egoire}}, \bibinfo {author} {\bibfnamefont
  {A.}~\bibnamefont {Kyanowski}}, \bibinfo {author} {\bibfnamefont
  {W.}~\bibnamefont {Mittig}}, \bibinfo {author} {\bibfnamefont
  {A.}~\bibnamefont {P\'eghaire}}, \bibinfo {author} {\bibfnamefont
  {J.}~\bibnamefont {P\'eter}}, \bibinfo {author} {\bibfnamefont
  {F.}~\bibnamefont {Saint-Laurent}}, \bibinfo {author} {\bibfnamefont {Y.~P.}\
  \bibnamefont {Viyogi}}, \bibinfo {author} {\bibfnamefont {B.}~\bibnamefont
  {Zwieglinski}}, \bibinfo {author} {\bibfnamefont {G.}~\bibnamefont {Bizard}},
  \bibinfo {author} {\bibfnamefont {F.}~\bibnamefont {Lef\`ebvres}}, \bibinfo
  {author} {\bibfnamefont {B.}~\bibnamefont {Tamain}}, \ and\ \bibinfo {author}
  {\bibfnamefont {J.}~\bibnamefont {Qu\'ebert}},\ }\href {\doibase
  10.1103/PhysRevLett.55.177} {\bibfield  {journal} {\bibinfo  {journal} {Phys.
  Rev. Lett.}\ }\textbf {\bibinfo {volume} {55}},\ \bibinfo {pages} {177}
  (\bibinfo {year} {1985})}\BibitemShut {NoStop}%
\bibitem [{\citenamefont {Pochodzalla}\ \emph {et~al.}(1987)\citenamefont
  {Pochodzalla}, \citenamefont {Gelbke}, \citenamefont {Lynch}, \citenamefont
  {Maier}, \citenamefont {Ardouin}, \citenamefont {Delagrange}, \citenamefont
  {Doubre}, \citenamefont {Gr\'egoire}, \citenamefont {Kyanowski},
  \citenamefont {Mittig}, \citenamefont {P\'eghaire}, \citenamefont {P\'eter},
  \citenamefont {Saint-Laurent}, \citenamefont {Zwieglinski}, \citenamefont
  {Bizard}, \citenamefont {Lef\`ebvres}, \citenamefont {Tamain}, \citenamefont
  {Qu\'ebert}, \citenamefont {Viyogi}, \citenamefont {Friedman},\ and\
  \citenamefont {Boal}}]{PhysRevC.35.1695}%
  \BibitemOpen
  \bibfield  {author} {\bibinfo {author} {\bibfnamefont {J.}~\bibnamefont
  {Pochodzalla}}, \bibinfo {author} {\bibfnamefont {C.~K.}\ \bibnamefont
  {Gelbke}}, \bibinfo {author} {\bibfnamefont {W.~G.}\ \bibnamefont {Lynch}},
  \bibinfo {author} {\bibfnamefont {M.}~\bibnamefont {Maier}}, \bibinfo
  {author} {\bibfnamefont {D.}~\bibnamefont {Ardouin}}, \bibinfo {author}
  {\bibfnamefont {H.}~\bibnamefont {Delagrange}}, \bibinfo {author}
  {\bibfnamefont {H.}~\bibnamefont {Doubre}}, \bibinfo {author} {\bibfnamefont
  {C.}~\bibnamefont {Gr\'egoire}}, \bibinfo {author} {\bibfnamefont
  {A.}~\bibnamefont {Kyanowski}}, \bibinfo {author} {\bibfnamefont
  {W.}~\bibnamefont {Mittig}}, \bibinfo {author} {\bibfnamefont
  {A.}~\bibnamefont {P\'eghaire}}, \bibinfo {author} {\bibfnamefont
  {J.}~\bibnamefont {P\'eter}}, \bibinfo {author} {\bibfnamefont
  {F.}~\bibnamefont {Saint-Laurent}}, \bibinfo {author} {\bibfnamefont
  {B.}~\bibnamefont {Zwieglinski}}, \bibinfo {author} {\bibfnamefont
  {G.}~\bibnamefont {Bizard}}, \bibinfo {author} {\bibfnamefont
  {F.}~\bibnamefont {Lef\`ebvres}}, \bibinfo {author} {\bibfnamefont
  {B.}~\bibnamefont {Tamain}}, \bibinfo {author} {\bibfnamefont
  {J.}~\bibnamefont {Qu\'ebert}}, \bibinfo {author} {\bibfnamefont {Y.~P.}\
  \bibnamefont {Viyogi}}, \bibinfo {author} {\bibfnamefont {W.~A.}\
  \bibnamefont {Friedman}}, \ and\ \bibinfo {author} {\bibfnamefont {D.~H.}\
  \bibnamefont {Boal}},\ }\href {\doibase 10.1103/PhysRevC.35.1695} {\bibfield
  {journal} {\bibinfo  {journal} {Phys. Rev. C}\ }\textbf {\bibinfo {volume}
  {35}},\ \bibinfo {pages} {1695} (\bibinfo {year} {1987})}\BibitemShut
  {NoStop}%
\bibitem [{\citenamefont {Wieland}\ \emph {et~al.}(2006)\citenamefont
  {Wieland}, \citenamefont {Bracco}, \citenamefont {Camera}, \citenamefont
  {Benzoni}, \citenamefont {Blasi}, \citenamefont {Brambilla}, \citenamefont
  {Crespi}, \citenamefont {Giussani}, \citenamefont {Leoni}, \citenamefont
  {Mason}, \citenamefont {Million}, \citenamefont {Moroni}, \citenamefont
  {Barlini}, \citenamefont {Kravchuk}, \citenamefont {Gramegna}, \citenamefont
  {Lanchais}, \citenamefont {Mastinu}, \citenamefont {Maj}, \citenamefont
  {Brekiesz}, \citenamefont {Kmiecik}, \citenamefont {Bruno}, \citenamefont
  {Geraci}, \citenamefont {Vannini}, \citenamefont {Casini}, \citenamefont
  {Chiari}, \citenamefont {Nannini}, \citenamefont {Ordine},\ and\
  \citenamefont {Ormand}}]{PhysRevLett.97.012501}%
  \BibitemOpen
  \bibfield  {author} {\bibinfo {author} {\bibfnamefont {O.}~\bibnamefont
  {Wieland}}, \bibinfo {author} {\bibfnamefont {A.}~\bibnamefont {Bracco}},
  \bibinfo {author} {\bibfnamefont {F.}~\bibnamefont {Camera}}, \bibinfo
  {author} {\bibfnamefont {G.}~\bibnamefont {Benzoni}}, \bibinfo {author}
  {\bibfnamefont {N.}~\bibnamefont {Blasi}}, \bibinfo {author} {\bibfnamefont
  {S.}~\bibnamefont {Brambilla}}, \bibinfo {author} {\bibfnamefont
  {F.}~\bibnamefont {Crespi}}, \bibinfo {author} {\bibfnamefont
  {A.}~\bibnamefont {Giussani}}, \bibinfo {author} {\bibfnamefont
  {S.}~\bibnamefont {Leoni}}, \bibinfo {author} {\bibfnamefont
  {P.}~\bibnamefont {Mason}}, \bibinfo {author} {\bibfnamefont
  {B.}~\bibnamefont {Million}}, \bibinfo {author} {\bibfnamefont
  {A.}~\bibnamefont {Moroni}}, \bibinfo {author} {\bibfnamefont
  {S.}~\bibnamefont {Barlini}}, \bibinfo {author} {\bibfnamefont {V.~L.}\
  \bibnamefont {Kravchuk}}, \bibinfo {author} {\bibfnamefont {F.}~\bibnamefont
  {Gramegna}}, \bibinfo {author} {\bibfnamefont {A.}~\bibnamefont {Lanchais}},
  \bibinfo {author} {\bibfnamefont {P.}~\bibnamefont {Mastinu}}, \bibinfo
  {author} {\bibfnamefont {A.}~\bibnamefont {Maj}}, \bibinfo {author}
  {\bibfnamefont {M.}~\bibnamefont {Brekiesz}}, \bibinfo {author}
  {\bibfnamefont {M.}~\bibnamefont {Kmiecik}}, \bibinfo {author} {\bibfnamefont
  {M.}~\bibnamefont {Bruno}}, \bibinfo {author} {\bibfnamefont
  {E.}~\bibnamefont {Geraci}}, \bibinfo {author} {\bibfnamefont
  {G.}~\bibnamefont {Vannini}}, \bibinfo {author} {\bibfnamefont
  {G.}~\bibnamefont {Casini}}, \bibinfo {author} {\bibfnamefont
  {M.}~\bibnamefont {Chiari}}, \bibinfo {author} {\bibfnamefont
  {A.}~\bibnamefont {Nannini}}, \bibinfo {author} {\bibfnamefont
  {A.}~\bibnamefont {Ordine}}, \ and\ \bibinfo {author} {\bibfnamefont
  {E.}~\bibnamefont {Ormand}},\ }\href {\doibase 10.1103/PhysRevLett.97.012501}
  {\bibfield  {journal} {\bibinfo  {journal} {Phys. Rev. Lett.}\ }\textbf
  {\bibinfo {volume} {97}},\ \bibinfo {pages} {012501} (\bibinfo {year}
  {2006})}\BibitemShut {NoStop}%
\bibitem [{\citenamefont {Snover}(1986)}]{snover1986giant}%
  \BibitemOpen
  \bibfield  {author} {\bibinfo {author} {\bibfnamefont {K.~A.}\ \bibnamefont
  {Snover}},\ }\href@noop {} {\bibfield  {journal} {\bibinfo  {journal} {Annual
  Review of Nuclear and Particle Science}\ }\textbf {\bibinfo {volume} {36}},\
  \bibinfo {pages} {545} (\bibinfo {year} {1986})}\BibitemShut {NoStop}%
\bibitem [{\citenamefont {Corsi}\ \emph {et~al.}(2011)\citenamefont {Corsi},
  \citenamefont {Wieland}, \citenamefont {Barlini}, \citenamefont {Bracco},
  \citenamefont {Camera}, \citenamefont {Kravchuk}, \citenamefont {Baiocco},
  \citenamefont {Bardelli}, \citenamefont {Benzoni}, \citenamefont {Bini},
  \citenamefont {Blasi}, \citenamefont {Brambilla}, \citenamefont {Bruno},
  \citenamefont {Casini}, \citenamefont {Ciemala}, \citenamefont {Cinausero},
  \citenamefont {Crespi}, \citenamefont {D'Agostino}, \citenamefont
  {Degerlier}, \citenamefont {Giaz}, \citenamefont {Gramegna}, \citenamefont
  {Kmiecik}, \citenamefont {Leoni}, \citenamefont {Maj}, \citenamefont
  {Marchi}, \citenamefont {Mazurek}, \citenamefont {Meczynski}, \citenamefont
  {Million}, \citenamefont {Montanari}, \citenamefont {Morelli}, \citenamefont
  {Myalski}, \citenamefont {Nannini}, \citenamefont {Nicolini}, \citenamefont
  {Pasquali}, \citenamefont {Poggi}, \citenamefont {Vandone},\ and\
  \citenamefont {Vannini}}]{PhysRevC.84.041304}%
  \BibitemOpen
  \bibfield  {author} {\bibinfo {author} {\bibfnamefont {A.}~\bibnamefont
  {Corsi}}, \bibinfo {author} {\bibfnamefont {O.}~\bibnamefont {Wieland}},
  \bibinfo {author} {\bibfnamefont {S.}~\bibnamefont {Barlini}}, \bibinfo
  {author} {\bibfnamefont {A.}~\bibnamefont {Bracco}}, \bibinfo {author}
  {\bibfnamefont {F.}~\bibnamefont {Camera}}, \bibinfo {author} {\bibfnamefont
  {V.~L.}\ \bibnamefont {Kravchuk}}, \bibinfo {author} {\bibfnamefont
  {G.}~\bibnamefont {Baiocco}}, \bibinfo {author} {\bibfnamefont
  {L.}~\bibnamefont {Bardelli}}, \bibinfo {author} {\bibfnamefont
  {G.}~\bibnamefont {Benzoni}}, \bibinfo {author} {\bibfnamefont
  {M.}~\bibnamefont {Bini}}, \bibinfo {author} {\bibfnamefont {N.}~\bibnamefont
  {Blasi}}, \bibinfo {author} {\bibfnamefont {S.}~\bibnamefont {Brambilla}},
  \bibinfo {author} {\bibfnamefont {M.}~\bibnamefont {Bruno}}, \bibinfo
  {author} {\bibfnamefont {G.}~\bibnamefont {Casini}}, \bibinfo {author}
  {\bibfnamefont {M.}~\bibnamefont {Ciemala}}, \bibinfo {author} {\bibfnamefont
  {M.}~\bibnamefont {Cinausero}}, \bibinfo {author} {\bibfnamefont {F.~C.~L.}\
  \bibnamefont {Crespi}}, \bibinfo {author} {\bibfnamefont {M.}~\bibnamefont
  {D'Agostino}}, \bibinfo {author} {\bibfnamefont {M.}~\bibnamefont
  {Degerlier}}, \bibinfo {author} {\bibfnamefont {A.}~\bibnamefont {Giaz}},
  \bibinfo {author} {\bibfnamefont {F.}~\bibnamefont {Gramegna}}, \bibinfo
  {author} {\bibfnamefont {M.}~\bibnamefont {Kmiecik}}, \bibinfo {author}
  {\bibfnamefont {S.}~\bibnamefont {Leoni}}, \bibinfo {author} {\bibfnamefont
  {A.}~\bibnamefont {Maj}}, \bibinfo {author} {\bibfnamefont {T.}~\bibnamefont
  {Marchi}}, \bibinfo {author} {\bibfnamefont {K.}~\bibnamefont {Mazurek}},
  \bibinfo {author} {\bibfnamefont {W.}~\bibnamefont {Meczynski}}, \bibinfo
  {author} {\bibfnamefont {B.}~\bibnamefont {Million}}, \bibinfo {author}
  {\bibfnamefont {D.}~\bibnamefont {Montanari}}, \bibinfo {author}
  {\bibfnamefont {L.}~\bibnamefont {Morelli}}, \bibinfo {author} {\bibfnamefont
  {S.}~\bibnamefont {Myalski}}, \bibinfo {author} {\bibfnamefont
  {A.}~\bibnamefont {Nannini}}, \bibinfo {author} {\bibfnamefont
  {R.}~\bibnamefont {Nicolini}}, \bibinfo {author} {\bibfnamefont
  {G.}~\bibnamefont {Pasquali}}, \bibinfo {author} {\bibfnamefont
  {G.}~\bibnamefont {Poggi}}, \bibinfo {author} {\bibfnamefont
  {V.}~\bibnamefont {Vandone}}, \ and\ \bibinfo {author} {\bibfnamefont
  {G.}~\bibnamefont {Vannini}},\ }\href {\doibase 10.1103/PhysRevC.84.041304}
  {\bibfield  {journal} {\bibinfo  {journal} {Phys. Rev. C}\ }\textbf {\bibinfo
  {volume} {84}},\ \bibinfo {pages} {041304} (\bibinfo {year}
  {2011})}\BibitemShut {NoStop}%
\bibitem [{\citenamefont {Gaardhøje}(1988)}]{GAARDHOJE1988261}%
  \BibitemOpen
  \bibfield  {author} {\bibinfo {author} {\bibfnamefont {J.~J.}\ \bibnamefont
  {Gaardhøje}},\ }\href {\doibase
  https://doi.org/10.1016/0375-9474(88)90268-0} {\bibfield  {journal} {\bibinfo
   {journal} {Nuclear Physics A}\ }\textbf {\bibinfo {volume} {488}},\ \bibinfo
  {pages} {261} (\bibinfo {year} {1988})}\BibitemShut {NoStop}%
\bibitem [{\citenamefont {Kicińska-Habior}\ \emph {et~al.}(1993)\citenamefont
  {Kicińska-Habior}, \citenamefont {Snover}, \citenamefont {Behr},
  \citenamefont {Gossett}, \citenamefont {Alhassid},\ and\ \citenamefont
  {Whelan}}]{KICINSKAHABIOR1993225}%
  \BibitemOpen
  \bibfield  {author} {\bibinfo {author} {\bibfnamefont {M.}~\bibnamefont
  {Kicińska-Habior}}, \bibinfo {author} {\bibfnamefont {K.}~\bibnamefont
  {Snover}}, \bibinfo {author} {\bibfnamefont {J.}~\bibnamefont {Behr}},
  \bibinfo {author} {\bibfnamefont {C.}~\bibnamefont {Gossett}}, \bibinfo
  {author} {\bibfnamefont {Y.}~\bibnamefont {Alhassid}}, \ and\ \bibinfo
  {author} {\bibfnamefont {N.}~\bibnamefont {Whelan}},\ }\href {\doibase
  https://doi.org/10.1016/0370-2693(93)91276-S} {\bibfield  {journal} {\bibinfo
   {journal} {Physics Letters B}\ }\textbf {\bibinfo {volume} {308}},\ \bibinfo
  {pages} {225} (\bibinfo {year} {1993})}\BibitemShut {NoStop}%
\bibitem [{\citenamefont {Suraud}\ \emph {et~al.}(1989)\citenamefont {Suraud},
  \citenamefont {Pi},\ and\ \citenamefont {Schuck}}]{SURAUD1989294}%
  \BibitemOpen
  \bibfield  {author} {\bibinfo {author} {\bibfnamefont {E.}~\bibnamefont
  {Suraud}}, \bibinfo {author} {\bibfnamefont {M.}~\bibnamefont {Pi}}, \ and\
  \bibinfo {author} {\bibfnamefont {P.}~\bibnamefont {Schuck}},\ }\href
  {\doibase https://doi.org/10.1016/0375-9474(89)90088-2} {\bibfield  {journal}
  {\bibinfo  {journal} {Nuclear Physics A}\ }\textbf {\bibinfo {volume}
  {492}},\ \bibinfo {pages} {294} (\bibinfo {year} {1989})}\BibitemShut
  {NoStop}%
\bibitem [{\citenamefont {Bethe}(1937)}]{RevModPhys.9.69}%
  \BibitemOpen
  \bibfield  {author} {\bibinfo {author} {\bibfnamefont {H.~A.}\ \bibnamefont
  {Bethe}},\ }\href {\doibase 10.1103/RevModPhys.9.69} {\bibfield  {journal}
  {\bibinfo  {journal} {Rev. Mod. Phys.}\ }\textbf {\bibinfo {volume} {9}},\
  \bibinfo {pages} {69} (\bibinfo {year} {1937})}\BibitemShut {NoStop}%
\bibitem [{\citenamefont {Bonche}\ \emph {et~al.}(1976)\citenamefont {Bonche},
  \citenamefont {Koonin},\ and\ \citenamefont {Negele}}]{PhysRevC.13.1226}%
  \BibitemOpen
  \bibfield  {author} {\bibinfo {author} {\bibfnamefont {P.}~\bibnamefont
  {Bonche}}, \bibinfo {author} {\bibfnamefont {S.}~\bibnamefont {Koonin}}, \
  and\ \bibinfo {author} {\bibfnamefont {J.~W.}\ \bibnamefont {Negele}},\
  }\href {\doibase 10.1103/PhysRevC.13.1226} {\bibfield  {journal} {\bibinfo
  {journal} {Phys. Rev. C}\ }\textbf {\bibinfo {volume} {13}},\ \bibinfo
  {pages} {1226} (\bibinfo {year} {1976})}\BibitemShut {NoStop}%
\bibitem [{\citenamefont {Simenel}\ and\ \citenamefont
  {Umar}(2018)}]{SIMENEL201819}%
  \BibitemOpen
  \bibfield  {author} {\bibinfo {author} {\bibfnamefont {C.}~\bibnamefont
  {Simenel}}\ and\ \bibinfo {author} {\bibfnamefont {A.}~\bibnamefont {Umar}},\
  }\href {\doibase https://doi.org/10.1016/j.ppnp.2018.07.002} {\bibfield
  {journal} {\bibinfo  {journal} {Progress in Particle and Nuclear Physics}\
  }\textbf {\bibinfo {volume} {103}},\ \bibinfo {pages} {19} (\bibinfo {year}
  {2018})}\BibitemShut {NoStop}%
\bibitem [{\citenamefont {Suraud}(1987)}]{SURAUD1987109}%
  \BibitemOpen
  \bibfield  {author} {\bibinfo {author} {\bibfnamefont {E.}~\bibnamefont
  {Suraud}},\ }\href {\doibase https://doi.org/10.1016/0375-9474(87)90382-4}
  {\bibfield  {journal} {\bibinfo  {journal} {Nuclear Physics A}\ }\textbf
  {\bibinfo {volume} {462}},\ \bibinfo {pages} {109} (\bibinfo {year}
  {1987})}\BibitemShut {NoStop}%
\bibitem [{\citenamefont {Levit}\ and\ \citenamefont
  {Bonche}(1985)}]{LEVIT1985426}%
  \BibitemOpen
  \bibfield  {author} {\bibinfo {author} {\bibfnamefont {S.}~\bibnamefont
  {Levit}}\ and\ \bibinfo {author} {\bibfnamefont {P.}~\bibnamefont {Bonche}},\
  }\href {\doibase https://doi.org/10.1016/S0375-9474(85)90099-5} {\bibfield
  {journal} {\bibinfo  {journal} {Nuclear Physics A}\ }\textbf {\bibinfo
  {volume} {437}},\ \bibinfo {pages} {426} (\bibinfo {year}
  {1985})}\BibitemShut {NoStop}%
\bibitem [{\citenamefont {Qu}\ and\ \citenamefont
  {Zhang}(2019)}]{PhysRevC.99.014314}%
  \BibitemOpen
  \bibfield  {author} {\bibinfo {author} {\bibfnamefont {X.~Y.}\ \bibnamefont
  {Qu}}\ and\ \bibinfo {author} {\bibfnamefont {Y.}~\bibnamefont {Zhang}},\
  }\href {\doibase 10.1103/PhysRevC.99.014314} {\bibfield  {journal} {\bibinfo
  {journal} {Phys. Rev. C}\ }\textbf {\bibinfo {volume} {99}},\ \bibinfo
  {pages} {014314} (\bibinfo {year} {2019})}\BibitemShut {NoStop}%
\bibitem [{\citenamefont {Sun}\ \emph {et~al.}(2014)\citenamefont {Sun},
  \citenamefont {Zhang}, \citenamefont {Zhang}, \citenamefont {Hu},\ and\
  \citenamefont {Meng}}]{PhysRevC.90.054321}%
  \BibitemOpen
  \bibfield  {author} {\bibinfo {author} {\bibfnamefont {T.~T.}\ \bibnamefont
  {Sun}}, \bibinfo {author} {\bibfnamefont {S.~Q.}\ \bibnamefont {Zhang}},
  \bibinfo {author} {\bibfnamefont {Y.}~\bibnamefont {Zhang}}, \bibinfo
  {author} {\bibfnamefont {J.~N.}\ \bibnamefont {Hu}}, \ and\ \bibinfo {author}
  {\bibfnamefont {J.}~\bibnamefont {Meng}},\ }\href {\doibase
  10.1103/PhysRevC.90.054321} {\bibfield  {journal} {\bibinfo  {journal} {Phys.
  Rev. C}\ }\textbf {\bibinfo {volume} {90}},\ \bibinfo {pages} {054321}
  (\bibinfo {year} {2014})}\BibitemShut {NoStop}%
\bibitem [{\citenamefont {Pei}\ \emph {et~al.}(2011)\citenamefont {Pei},
  \citenamefont {Kruppa},\ and\ \citenamefont
  {Nazarewicz}}]{PhysRevC.84.024311}%
  \BibitemOpen
  \bibfield  {author} {\bibinfo {author} {\bibfnamefont {J.~C.}\ \bibnamefont
  {Pei}}, \bibinfo {author} {\bibfnamefont {A.~T.}\ \bibnamefont {Kruppa}}, \
  and\ \bibinfo {author} {\bibfnamefont {W.}~\bibnamefont {Nazarewicz}},\
  }\href {\doibase 10.1103/PhysRevC.84.024311} {\bibfield  {journal} {\bibinfo
  {journal} {Phys. Rev. C}\ }\textbf {\bibinfo {volume} {84}},\ \bibinfo
  {pages} {024311} (\bibinfo {year} {2011})}\BibitemShut {NoStop}%
\bibitem [{\citenamefont {Bonche}\ \emph {et~al.}(1985)\citenamefont {Bonche},
  \citenamefont {Levit},\ and\ \citenamefont {Vautherin}}]{BONCHE1985265}%
  \BibitemOpen
  \bibfield  {author} {\bibinfo {author} {\bibfnamefont {P.}~\bibnamefont
  {Bonche}}, \bibinfo {author} {\bibfnamefont {S.}~\bibnamefont {Levit}}, \
  and\ \bibinfo {author} {\bibfnamefont {D.}~\bibnamefont {Vautherin}},\ }\href
  {\doibase https://doi.org/10.1016/0375-9474(85)90199-X} {\bibfield  {journal}
  {\bibinfo  {journal} {Nuclear Physics A}\ }\textbf {\bibinfo {volume}
  {436}},\ \bibinfo {pages} {265} (\bibinfo {year} {1985})}\BibitemShut
  {NoStop}%
\bibitem [{\citenamefont {Bonche}\ \emph {et~al.}(1984)\citenamefont {Bonche},
  \citenamefont {Levit},\ and\ \citenamefont {Vautherin}}]{BONCHE1984278}%
  \BibitemOpen
  \bibfield  {author} {\bibinfo {author} {\bibfnamefont {P.}~\bibnamefont
  {Bonche}}, \bibinfo {author} {\bibfnamefont {S.}~\bibnamefont {Levit}}, \
  and\ \bibinfo {author} {\bibfnamefont {D.}~\bibnamefont {Vautherin}},\ }\href
  {\doibase https://doi.org/10.1016/0375-9474(84)90086-1} {\bibfield  {journal}
  {\bibinfo  {journal} {Nuclear Physics A}\ }\textbf {\bibinfo {volume}
  {427}},\ \bibinfo {pages} {278} (\bibinfo {year} {1984})}\BibitemShut
  {NoStop}%
\bibitem [{\citenamefont {Vertse}\ \emph {et~al.}(2000)\citenamefont {Vertse},
  \citenamefont {Kruppa},\ and\ \citenamefont
  {Nazarewicz}}]{PhysRevC.61.064317}%
  \BibitemOpen
  \bibfield  {author} {\bibinfo {author} {\bibfnamefont {T.}~\bibnamefont
  {Vertse}}, \bibinfo {author} {\bibfnamefont {A.~T.}\ \bibnamefont {Kruppa}},
  \ and\ \bibinfo {author} {\bibfnamefont {W.}~\bibnamefont {Nazarewicz}},\
  }\href {\doibase 10.1103/PhysRevC.61.064317} {\bibfield  {journal} {\bibinfo
  {journal} {Phys. Rev. C}\ }\textbf {\bibinfo {volume} {61}},\ \bibinfo
  {pages} {064317} (\bibinfo {year} {2000})}\BibitemShut {NoStop}%
\bibitem [{\citenamefont {Goodman}(1981{\natexlab{a}})}]{GOODMAN198130}%
  \BibitemOpen
  \bibfield  {author} {\bibinfo {author} {\bibfnamefont {A.~L.}\ \bibnamefont
  {Goodman}},\ }\href {\doibase https://doi.org/10.1016/0375-9474(81)90557-1}
  {\bibfield  {journal} {\bibinfo  {journal} {Nuclear Physics A}\ }\textbf
  {\bibinfo {volume} {352}},\ \bibinfo {pages} {30} (\bibinfo {year}
  {1981}{\natexlab{a}})}\BibitemShut {NoStop}%
\bibitem [{\citenamefont {Goodman}(1981{\natexlab{b}})}]{GOODMAN198145}%
  \BibitemOpen
  \bibfield  {author} {\bibinfo {author} {\bibfnamefont {A.~L.}\ \bibnamefont
  {Goodman}},\ }\href {\doibase https://doi.org/10.1016/0375-9474(81)90558-3}
  {\bibfield  {journal} {\bibinfo  {journal} {Nuclear Physics A}\ }\textbf
  {\bibinfo {volume} {352}},\ \bibinfo {pages} {45} (\bibinfo {year}
  {1981}{\natexlab{b}})}\BibitemShut {NoStop}%
\bibitem [{\citenamefont {Goodman}(1986)}]{PhysRevC.34.1942}%
  \BibitemOpen
  \bibfield  {author} {\bibinfo {author} {\bibfnamefont {A.~L.}\ \bibnamefont
  {Goodman}},\ }\href {\doibase 10.1103/PhysRevC.34.1942} {\bibfield  {journal}
  {\bibinfo  {journal} {Phys. Rev. C}\ }\textbf {\bibinfo {volume} {34}},\
  \bibinfo {pages} {1942} (\bibinfo {year} {1986})}\BibitemShut {NoStop}%
\bibitem [{\citenamefont {Egido}\ \emph {et~al.}(1986)\citenamefont {Egido},
  \citenamefont {Ring},\ and\ \citenamefont {Mang}}]{EGIDO198677}%
  \BibitemOpen
  \bibfield  {author} {\bibinfo {author} {\bibfnamefont {J.}~\bibnamefont
  {Egido}}, \bibinfo {author} {\bibfnamefont {P.}~\bibnamefont {Ring}}, \ and\
  \bibinfo {author} {\bibfnamefont {H.}~\bibnamefont {Mang}},\ }\href {\doibase
  https://doi.org/10.1016/0375-9474(86)90242-3} {\bibfield  {journal} {\bibinfo
   {journal} {Nuclear Physics A}\ }\textbf {\bibinfo {volume} {451}},\ \bibinfo
  {pages} {77} (\bibinfo {year} {1986})}\BibitemShut {NoStop}%
\bibitem [{\citenamefont {Erler}\ \emph {et~al.}(2012)\citenamefont {Erler},
  \citenamefont {Birge}, \citenamefont {Kortelainen}, \citenamefont
  {Nazarewicz}, \citenamefont {Olsen}, \citenamefont {Perhac},\ and\
  \citenamefont {Stoitsov}}]{erler2012limits}%
  \BibitemOpen
  \bibfield  {author} {\bibinfo {author} {\bibfnamefont {J.}~\bibnamefont
  {Erler}}, \bibinfo {author} {\bibfnamefont {N.}~\bibnamefont {Birge}},
  \bibinfo {author} {\bibfnamefont {M.}~\bibnamefont {Kortelainen}}, \bibinfo
  {author} {\bibfnamefont {W.}~\bibnamefont {Nazarewicz}}, \bibinfo {author}
  {\bibfnamefont {E.}~\bibnamefont {Olsen}}, \bibinfo {author} {\bibfnamefont
  {A.~M.}\ \bibnamefont {Perhac}}, \ and\ \bibinfo {author} {\bibfnamefont
  {M.}~\bibnamefont {Stoitsov}},\ }\href@noop {} {\bibfield  {journal}
  {\bibinfo  {journal} {Nature}\ }\textbf {\bibinfo {volume} {486}},\ \bibinfo
  {pages} {509} (\bibinfo {year} {2012})}\BibitemShut {NoStop}%
\bibitem [{\citenamefont {Goriely}\ \emph {et~al.}(2009)\citenamefont
  {Goriely}, \citenamefont {Hilaire}, \citenamefont {Girod},\ and\
  \citenamefont {P\'eru}}]{PhysRevLett.102.242501}%
  \BibitemOpen
  \bibfield  {author} {\bibinfo {author} {\bibfnamefont {S.}~\bibnamefont
  {Goriely}}, \bibinfo {author} {\bibfnamefont {S.}~\bibnamefont {Hilaire}},
  \bibinfo {author} {\bibfnamefont {M.}~\bibnamefont {Girod}}, \ and\ \bibinfo
  {author} {\bibfnamefont {S.}~\bibnamefont {P\'eru}},\ }\href {\doibase
  10.1103/PhysRevLett.102.242501} {\bibfield  {journal} {\bibinfo  {journal}
  {Phys. Rev. Lett.}\ }\textbf {\bibinfo {volume} {102}},\ \bibinfo {pages}
  {242501} (\bibinfo {year} {2009})}\BibitemShut {NoStop}%
\bibitem [{\citenamefont {Delaroche}\ \emph {et~al.}(2010)\citenamefont
  {Delaroche}, \citenamefont {Girod}, \citenamefont {Libert}, \citenamefont
  {Goutte}, \citenamefont {Hilaire}, \citenamefont {P\'eru}, \citenamefont
  {Pillet},\ and\ \citenamefont {Bertsch}}]{PhysRevC.81.014303}%
  \BibitemOpen
  \bibfield  {author} {\bibinfo {author} {\bibfnamefont {J.~P.}\ \bibnamefont
  {Delaroche}}, \bibinfo {author} {\bibfnamefont {M.}~\bibnamefont {Girod}},
  \bibinfo {author} {\bibfnamefont {J.}~\bibnamefont {Libert}}, \bibinfo
  {author} {\bibfnamefont {H.}~\bibnamefont {Goutte}}, \bibinfo {author}
  {\bibfnamefont {S.}~\bibnamefont {Hilaire}}, \bibinfo {author} {\bibfnamefont
  {S.}~\bibnamefont {P\'eru}}, \bibinfo {author} {\bibfnamefont
  {N.}~\bibnamefont {Pillet}}, \ and\ \bibinfo {author} {\bibfnamefont {G.~F.}\
  \bibnamefont {Bertsch}},\ }\href {\doibase 10.1103/PhysRevC.81.014303}
  {\bibfield  {journal} {\bibinfo  {journal} {Phys. Rev. C}\ }\textbf {\bibinfo
  {volume} {81}},\ \bibinfo {pages} {014303} (\bibinfo {year}
  {2010})}\BibitemShut {NoStop}%
\bibitem [{\citenamefont {Y{\"u}ksel}\ \emph {et~al.}(2014)\citenamefont
  {Y{\"u}ksel}, \citenamefont {Khan}, \citenamefont {Bozkurt},\ and\
  \citenamefont {Col{\`o}}}]{Yuksel2014}%
  \BibitemOpen
  \bibfield  {author} {\bibinfo {author} {\bibfnamefont {E.}~\bibnamefont
  {Y{\"u}ksel}}, \bibinfo {author} {\bibfnamefont {E.}~\bibnamefont {Khan}},
  \bibinfo {author} {\bibfnamefont {K.}~\bibnamefont {Bozkurt}}, \ and\
  \bibinfo {author} {\bibfnamefont {G.}~\bibnamefont {Col{\`o}}},\ }\href
  {\doibase 10.1140/epja/i2014-14160-4} {\bibfield  {journal} {\bibinfo
  {journal} {The European Physical Journal A}\ }\textbf {\bibinfo {volume}
  {50}},\ \bibinfo {pages} {160} (\bibinfo {year} {2014})}\BibitemShut
  {NoStop}%
\bibitem [{\citenamefont {Niu}\ \emph {et~al.}(2013)\citenamefont {Niu},
  \citenamefont {Niu}, \citenamefont {Paar}, \citenamefont {Vretenar},
  \citenamefont {Wang}, \citenamefont {Bai},\ and\ \citenamefont
  {Meng}}]{PhysRevC.88.034308}%
  \BibitemOpen
  \bibfield  {author} {\bibinfo {author} {\bibfnamefont {Y.~F.}\ \bibnamefont
  {Niu}}, \bibinfo {author} {\bibfnamefont {Z.~M.}\ \bibnamefont {Niu}},
  \bibinfo {author} {\bibfnamefont {N.}~\bibnamefont {Paar}}, \bibinfo {author}
  {\bibfnamefont {D.}~\bibnamefont {Vretenar}}, \bibinfo {author}
  {\bibfnamefont {G.~H.}\ \bibnamefont {Wang}}, \bibinfo {author}
  {\bibfnamefont {J.~S.}\ \bibnamefont {Bai}}, \ and\ \bibinfo {author}
  {\bibfnamefont {J.}~\bibnamefont {Meng}},\ }\href {\doibase
  10.1103/PhysRevC.88.034308} {\bibfield  {journal} {\bibinfo  {journal} {Phys.
  Rev. C}\ }\textbf {\bibinfo {volume} {88}},\ \bibinfo {pages} {034308}
  (\bibinfo {year} {2013})}\BibitemShut {NoStop}%
\bibitem [{\citenamefont {Li}\ \emph {et~al.}(2015)\citenamefont {Li},
  \citenamefont {Margueron}, \citenamefont {Long},\ and\ \citenamefont
  {Van~Giai}}]{PhysRevC.92.014302}%
  \BibitemOpen
  \bibfield  {author} {\bibinfo {author} {\bibfnamefont {J.~J.}\ \bibnamefont
  {Li}}, \bibinfo {author} {\bibfnamefont {J.}~\bibnamefont {Margueron}},
  \bibinfo {author} {\bibfnamefont {W.~H.}\ \bibnamefont {Long}}, \ and\
  \bibinfo {author} {\bibfnamefont {N.}~\bibnamefont {Van~Giai}},\ }\href
  {\doibase 10.1103/PhysRevC.92.014302} {\bibfield  {journal} {\bibinfo
  {journal} {Phys. Rev. C}\ }\textbf {\bibinfo {volume} {92}},\ \bibinfo
  {pages} {014302} (\bibinfo {year} {2015})}\BibitemShut {NoStop}%
\bibitem [{\citenamefont {Schunck}\ \emph {et~al.}(2015)\citenamefont
  {Schunck}, \citenamefont {Duke},\ and\ \citenamefont
  {Carr}}]{PhysRevC.91.034327}%
  \BibitemOpen
  \bibfield  {author} {\bibinfo {author} {\bibfnamefont {N.}~\bibnamefont
  {Schunck}}, \bibinfo {author} {\bibfnamefont {D.}~\bibnamefont {Duke}}, \
  and\ \bibinfo {author} {\bibfnamefont {H.}~\bibnamefont {Carr}},\ }\href
  {\doibase 10.1103/PhysRevC.91.034327} {\bibfield  {journal} {\bibinfo
  {journal} {Phys. Rev. C}\ }\textbf {\bibinfo {volume} {91}},\ \bibinfo
  {pages} {034327} (\bibinfo {year} {2015})}\BibitemShut {NoStop}%
\bibitem [{\citenamefont {Zhu}\ and\ \citenamefont
  {Pei}(2016)}]{PhysRevC.94.024329}%
  \BibitemOpen
  \bibfield  {author} {\bibinfo {author} {\bibfnamefont {Y.}~\bibnamefont
  {Zhu}}\ and\ \bibinfo {author} {\bibfnamefont {J.~C.}\ \bibnamefont {Pei}},\
  }\href {\doibase 10.1103/PhysRevC.94.024329} {\bibfield  {journal} {\bibinfo
  {journal} {Phys. Rev. C}\ }\textbf {\bibinfo {volume} {94}},\ \bibinfo
  {pages} {024329} (\bibinfo {year} {2016})}\BibitemShut {NoStop}%
\bibitem [{\citenamefont {Y\"uksel}\ \emph {et~al.}(2022)\citenamefont
  {Y\"uksel}, \citenamefont {Mercier}, \citenamefont {Ebran},\ and\
  \citenamefont {Khan}}]{PhysRevC.106.054309}%
  \BibitemOpen
  \bibfield  {author} {\bibinfo {author} {\bibfnamefont {E.}~\bibnamefont
  {Y\"uksel}}, \bibinfo {author} {\bibfnamefont {F.}~\bibnamefont {Mercier}},
  \bibinfo {author} {\bibfnamefont {J.-P.}\ \bibnamefont {Ebran}}, \ and\
  \bibinfo {author} {\bibfnamefont {E.}~\bibnamefont {Khan}},\ }\href {\doibase
  10.1103/PhysRevC.106.054309} {\bibfield  {journal} {\bibinfo  {journal}
  {Phys. Rev. C}\ }\textbf {\bibinfo {volume} {106}},\ \bibinfo {pages}
  {054309} (\bibinfo {year} {2022})}\BibitemShut {NoStop}%
\bibitem [{\citenamefont {Zhu}\ and\ \citenamefont
  {Pei}(2014)}]{PhysRevC.90.054316}%
  \BibitemOpen
  \bibfield  {author} {\bibinfo {author} {\bibfnamefont {Y.}~\bibnamefont
  {Zhu}}\ and\ \bibinfo {author} {\bibfnamefont {J.~C.}\ \bibnamefont {Pei}},\
  }\href {\doibase 10.1103/PhysRevC.90.054316} {\bibfield  {journal} {\bibinfo
  {journal} {Phys. Rev. C}\ }\textbf {\bibinfo {volume} {90}},\ \bibinfo
  {pages} {054316} (\bibinfo {year} {2014})}\BibitemShut {NoStop}%
\bibitem [{\citenamefont {Lisboa}\ \emph {et~al.}(2010)\citenamefont {Lisboa},
  \citenamefont {Malheiro},\ and\ \citenamefont {Carlson}}]{LISBOA2010345}%
  \BibitemOpen
  \bibfield  {author} {\bibinfo {author} {\bibfnamefont {R.}~\bibnamefont
  {Lisboa}}, \bibinfo {author} {\bibfnamefont {M.}~\bibnamefont {Malheiro}}, \
  and\ \bibinfo {author} {\bibfnamefont {B.}~\bibnamefont {Carlson}},\ }\href
  {\doibase https://doi.org/10.1016/j.nuclphysbps.2010.02.057} {\bibfield
  {journal} {\bibinfo  {journal} {Nuclear Physics B - Proceedings Supplements}\
  }\textbf {\bibinfo {volume} {199}},\ \bibinfo {pages} {345} (\bibinfo {year}
  {2010})},\ \bibinfo {note} {proceedings of the International Workshop Light
  Cone 2009 (LC2009): Relativistic Hadronic and Particle Physics}\BibitemShut
  {NoStop}%
\bibitem [{\citenamefont {Yüksel}(2021)}]{YUKSEL2021122238}%
  \BibitemOpen
  \bibfield  {author} {\bibinfo {author} {\bibfnamefont {E.}~\bibnamefont
  {Yüksel}},\ }\href {\doibase
  https://doi.org/10.1016/j.nuclphysa.2021.122238} {\bibfield  {journal}
  {\bibinfo  {journal} {Nuclear Physics A}\ }\textbf {\bibinfo {volume}
  {1014}},\ \bibinfo {pages} {122238} (\bibinfo {year} {2021})}\BibitemShut
  {NoStop}%
\bibitem [{\citenamefont {Vretenar}\ \emph {et~al.}(2005)\citenamefont
  {Vretenar}, \citenamefont {Afanasjev}, \citenamefont {Lalazissis},\ and\
  \citenamefont {Ring}}]{VRETENAR2005101}%
  \BibitemOpen
  \bibfield  {author} {\bibinfo {author} {\bibfnamefont {D.}~\bibnamefont
  {Vretenar}}, \bibinfo {author} {\bibfnamefont {A.}~\bibnamefont {Afanasjev}},
  \bibinfo {author} {\bibfnamefont {G.}~\bibnamefont {Lalazissis}}, \ and\
  \bibinfo {author} {\bibfnamefont {P.}~\bibnamefont {Ring}},\ }\href {\doibase
  https://doi.org/10.1016/j.physrep.2004.10.001} {\bibfield  {journal}
  {\bibinfo  {journal} {Physics Reports}\ }\textbf {\bibinfo {volume} {409}},\
  \bibinfo {pages} {101} (\bibinfo {year} {2005})}\BibitemShut {NoStop}%
\bibitem [{\citenamefont {Nikšić}\ \emph {et~al.}(2011)\citenamefont
  {Nikšić}, \citenamefont {Vretenar},\ and\ \citenamefont
  {Ring}}]{NIKSIC2011519}%
  \BibitemOpen
  \bibfield  {author} {\bibinfo {author} {\bibfnamefont {T.}~\bibnamefont
  {Nikšić}}, \bibinfo {author} {\bibfnamefont {D.}~\bibnamefont {Vretenar}},
  \ and\ \bibinfo {author} {\bibfnamefont {P.}~\bibnamefont {Ring}},\ }\href
  {\doibase https://doi.org/10.1016/j.ppnp.2011.01.055} {\bibfield  {journal}
  {\bibinfo  {journal} {Progress in Particle and Nuclear Physics}\ }\textbf
  {\bibinfo {volume} {66}},\ \bibinfo {pages} {519} (\bibinfo {year}
  {2011})}\BibitemShut {NoStop}%
\bibitem [{\citenamefont {Lalazissis}\ \emph {et~al.}(2005)\citenamefont
  {Lalazissis}, \citenamefont {Nik\ifmmode \check{s}\else
  \v{s}\fi{}i\ifmmode~\acute{c}\else \'{c}\fi{}}, \citenamefont {Vretenar},\
  and\ \citenamefont {Ring}}]{PhysRevC.71.024312}%
  \BibitemOpen
  \bibfield  {author} {\bibinfo {author} {\bibfnamefont {G.~A.}\ \bibnamefont
  {Lalazissis}}, \bibinfo {author} {\bibfnamefont {T.}~\bibnamefont
  {Nik\ifmmode \check{s}\else \v{s}\fi{}i\ifmmode~\acute{c}\else \'{c}\fi{}}},
  \bibinfo {author} {\bibfnamefont {D.}~\bibnamefont {Vretenar}}, \ and\
  \bibinfo {author} {\bibfnamefont {P.}~\bibnamefont {Ring}},\ }\href {\doibase
  10.1103/PhysRevC.71.024312} {\bibfield  {journal} {\bibinfo  {journal} {Phys.
  Rev. C}\ }\textbf {\bibinfo {volume} {71}},\ \bibinfo {pages} {024312}
  (\bibinfo {year} {2005})}\BibitemShut {NoStop}%
\bibitem [{\citenamefont {Nik\ifmmode \check{s}\else
  \v{s}\fi{}i\ifmmode~\acute{c}\else \'{c}\fi{}}\ \emph
  {et~al.}(2008)\citenamefont {Nik\ifmmode \check{s}\else
  \v{s}\fi{}i\ifmmode~\acute{c}\else \'{c}\fi{}}, \citenamefont {Vretenar},\
  and\ \citenamefont {Ring}}]{PhysRevC.78.034318}%
  \BibitemOpen
  \bibfield  {author} {\bibinfo {author} {\bibfnamefont {T.}~\bibnamefont
  {Nik\ifmmode \check{s}\else \v{s}\fi{}i\ifmmode~\acute{c}\else \'{c}\fi{}}},
  \bibinfo {author} {\bibfnamefont {D.}~\bibnamefont {Vretenar}}, \ and\
  \bibinfo {author} {\bibfnamefont {P.}~\bibnamefont {Ring}},\ }\href {\doibase
  10.1103/PhysRevC.78.034318} {\bibfield  {journal} {\bibinfo  {journal} {Phys.
  Rev. C}\ }\textbf {\bibinfo {volume} {78}},\ \bibinfo {pages} {034318}
  (\bibinfo {year} {2008})}\BibitemShut {NoStop}%
\bibitem [{\citenamefont {Afanasjev}\ \emph {et~al.}(2013)\citenamefont
  {Afanasjev}, \citenamefont {Agbemava}, \citenamefont {Ray},\ and\
  \citenamefont {Ring}}]{AFANASJEV2013680}%
  \BibitemOpen
  \bibfield  {author} {\bibinfo {author} {\bibfnamefont {A.}~\bibnamefont
  {Afanasjev}}, \bibinfo {author} {\bibfnamefont {S.}~\bibnamefont {Agbemava}},
  \bibinfo {author} {\bibfnamefont {D.}~\bibnamefont {Ray}}, \ and\ \bibinfo
  {author} {\bibfnamefont {P.}~\bibnamefont {Ring}},\ }\href {\doibase
  https://doi.org/10.1016/j.physletb.2013.09.017} {\bibfield  {journal}
  {\bibinfo  {journal} {Physics Letters B}\ }\textbf {\bibinfo {volume}
  {726}},\ \bibinfo {pages} {680} (\bibinfo {year} {2013})}\BibitemShut
  {NoStop}%
\bibitem [{\citenamefont {Agbemava}\ \emph {et~al.}(2014)\citenamefont
  {Agbemava}, \citenamefont {Afanasjev}, \citenamefont {Ray},\ and\
  \citenamefont {Ring}}]{PhysRevC.89.054320}%
  \BibitemOpen
  \bibfield  {author} {\bibinfo {author} {\bibfnamefont {S.~E.}\ \bibnamefont
  {Agbemava}}, \bibinfo {author} {\bibfnamefont {A.~V.}\ \bibnamefont
  {Afanasjev}}, \bibinfo {author} {\bibfnamefont {D.}~\bibnamefont {Ray}}, \
  and\ \bibinfo {author} {\bibfnamefont {P.}~\bibnamefont {Ring}},\ }\href
  {\doibase 10.1103/PhysRevC.89.054320} {\bibfield  {journal} {\bibinfo
  {journal} {Phys. Rev. C}\ }\textbf {\bibinfo {volume} {89}},\ \bibinfo
  {pages} {054320} (\bibinfo {year} {2014})}\BibitemShut {NoStop}%
\bibitem [{\citenamefont {Zhang}\ \emph {et~al.}(2022)\citenamefont {Zhang},
  \citenamefont {Cheoun}, \citenamefont {Choi}, \citenamefont {Chong},
  \citenamefont {Dong}, \citenamefont {Dong}, \citenamefont {Du}, \citenamefont
  {Geng}, \citenamefont {Ha}, \citenamefont {He}, \citenamefont {Heo},
  \citenamefont {Ho}, \citenamefont {In}, \citenamefont {Kim}, \citenamefont
  {Kim}, \citenamefont {Lee}, \citenamefont {Lee}, \citenamefont {Li},
  \citenamefont {Li}, \citenamefont {Luo}, \citenamefont {Meng}, \citenamefont
  {Mun}, \citenamefont {Niu}, \citenamefont {Pan}, \citenamefont
  {Papakonstantinou}, \citenamefont {Shang}, \citenamefont {Shen},
  \citenamefont {Shen}, \citenamefont {Sun}, \citenamefont {Sun}, \citenamefont
  {Tam}, \citenamefont {Thaivayongnou}, \citenamefont {Wang}, \citenamefont
  {Wang}, \citenamefont {Wong}, \citenamefont {Wu}, \citenamefont {Wu},
  \citenamefont {Xia}, \citenamefont {Yan}, \citenamefont {Yeung},
  \citenamefont {Yiu}, \citenamefont {Zhang}, \citenamefont {Zhang},
  \citenamefont {Zhang}, \citenamefont {Zhao},\ and\ \citenamefont
  {Zhou}}]{ZHANG2022101488}%
  \BibitemOpen
  \bibfield  {author} {\bibinfo {author} {\bibfnamefont {K.}~\bibnamefont
  {Zhang}}, \bibinfo {author} {\bibfnamefont {M.-K.}\ \bibnamefont {Cheoun}},
  \bibinfo {author} {\bibfnamefont {Y.-B.}\ \bibnamefont {Choi}}, \bibinfo
  {author} {\bibfnamefont {P.~S.}\ \bibnamefont {Chong}}, \bibinfo {author}
  {\bibfnamefont {J.}~\bibnamefont {Dong}}, \bibinfo {author} {\bibfnamefont
  {Z.}~\bibnamefont {Dong}}, \bibinfo {author} {\bibfnamefont {X.}~\bibnamefont
  {Du}}, \bibinfo {author} {\bibfnamefont {L.}~\bibnamefont {Geng}}, \bibinfo
  {author} {\bibfnamefont {E.}~\bibnamefont {Ha}}, \bibinfo {author}
  {\bibfnamefont {X.-T.}\ \bibnamefont {He}}, \bibinfo {author} {\bibfnamefont
  {C.}~\bibnamefont {Heo}}, \bibinfo {author} {\bibfnamefont {M.~C.}\
  \bibnamefont {Ho}}, \bibinfo {author} {\bibfnamefont {E.~J.}\ \bibnamefont
  {In}}, \bibinfo {author} {\bibfnamefont {S.}~\bibnamefont {Kim}}, \bibinfo
  {author} {\bibfnamefont {Y.}~\bibnamefont {Kim}}, \bibinfo {author}
  {\bibfnamefont {C.-H.}\ \bibnamefont {Lee}}, \bibinfo {author} {\bibfnamefont
  {J.}~\bibnamefont {Lee}}, \bibinfo {author} {\bibfnamefont {H.}~\bibnamefont
  {Li}}, \bibinfo {author} {\bibfnamefont {Z.}~\bibnamefont {Li}}, \bibinfo
  {author} {\bibfnamefont {T.}~\bibnamefont {Luo}}, \bibinfo {author}
  {\bibfnamefont {J.}~\bibnamefont {Meng}}, \bibinfo {author} {\bibfnamefont
  {M.-H.}\ \bibnamefont {Mun}}, \bibinfo {author} {\bibfnamefont
  {Z.}~\bibnamefont {Niu}}, \bibinfo {author} {\bibfnamefont {C.}~\bibnamefont
  {Pan}}, \bibinfo {author} {\bibfnamefont {P.}~\bibnamefont
  {Papakonstantinou}}, \bibinfo {author} {\bibfnamefont {X.}~\bibnamefont
  {Shang}}, \bibinfo {author} {\bibfnamefont {C.}~\bibnamefont {Shen}},
  \bibinfo {author} {\bibfnamefont {G.}~\bibnamefont {Shen}}, \bibinfo {author}
  {\bibfnamefont {W.}~\bibnamefont {Sun}}, \bibinfo {author} {\bibfnamefont
  {X.-X.}\ \bibnamefont {Sun}}, \bibinfo {author} {\bibfnamefont {C.~K.}\
  \bibnamefont {Tam}}, \bibinfo {author} {\bibnamefont {Thaivayongnou}},
  \bibinfo {author} {\bibfnamefont {C.}~\bibnamefont {Wang}}, \bibinfo {author}
  {\bibfnamefont {X.}~\bibnamefont {Wang}}, \bibinfo {author} {\bibfnamefont
  {S.~H.}\ \bibnamefont {Wong}}, \bibinfo {author} {\bibfnamefont
  {J.}~\bibnamefont {Wu}}, \bibinfo {author} {\bibfnamefont {X.}~\bibnamefont
  {Wu}}, \bibinfo {author} {\bibfnamefont {X.}~\bibnamefont {Xia}}, \bibinfo
  {author} {\bibfnamefont {Y.}~\bibnamefont {Yan}}, \bibinfo {author}
  {\bibfnamefont {R.~W.-Y.}\ \bibnamefont {Yeung}}, \bibinfo {author}
  {\bibfnamefont {T.~C.}\ \bibnamefont {Yiu}}, \bibinfo {author} {\bibfnamefont
  {S.}~\bibnamefont {Zhang}}, \bibinfo {author} {\bibfnamefont
  {W.}~\bibnamefont {Zhang}}, \bibinfo {author} {\bibfnamefont
  {X.}~\bibnamefont {Zhang}}, \bibinfo {author} {\bibfnamefont
  {Q.}~\bibnamefont {Zhao}}, \ and\ \bibinfo {author} {\bibfnamefont {S.-G.}\
  \bibnamefont {Zhou}},\ }\href {\doibase
  https://doi.org/10.1016/j.adt.2022.101488} {\bibfield  {journal} {\bibinfo
  {journal} {Atomic Data and Nuclear Data Tables}\ }\textbf {\bibinfo {volume}
  {144}},\ \bibinfo {pages} {101488} (\bibinfo {year} {2022})}\BibitemShut
  {NoStop}%
\bibitem [{\citenamefont {Belabbas}\ \emph {et~al.}(2017)\citenamefont
  {Belabbas}, \citenamefont {Li},\ and\ \citenamefont
  {Margueron}}]{PhysRevC.96.024304}%
  \BibitemOpen
  \bibfield  {author} {\bibinfo {author} {\bibfnamefont {M.}~\bibnamefont
  {Belabbas}}, \bibinfo {author} {\bibfnamefont {J.~J.}\ \bibnamefont {Li}}, \
  and\ \bibinfo {author} {\bibfnamefont {J.}~\bibnamefont {Margueron}},\ }\href
  {\doibase 10.1103/PhysRevC.96.024304} {\bibfield  {journal} {\bibinfo
  {journal} {Phys. Rev. C}\ }\textbf {\bibinfo {volume} {96}},\ \bibinfo
  {pages} {024304} (\bibinfo {year} {2017})}\BibitemShut {NoStop}%
\bibitem [{\citenamefont {Agrawal}\ \emph {et~al.}(2000)\citenamefont
  {Agrawal}, \citenamefont {Sil}, \citenamefont {De},\ and\ \citenamefont
  {Samaddar}}]{PhysRevC.62.044307}%
  \BibitemOpen
  \bibfield  {author} {\bibinfo {author} {\bibfnamefont {B.~K.}\ \bibnamefont
  {Agrawal}}, \bibinfo {author} {\bibfnamefont {T.}~\bibnamefont {Sil}},
  \bibinfo {author} {\bibfnamefont {J.~N.}\ \bibnamefont {De}}, \ and\ \bibinfo
  {author} {\bibfnamefont {S.~K.}\ \bibnamefont {Samaddar}},\ }\href {\doibase
  10.1103/PhysRevC.62.044307} {\bibfield  {journal} {\bibinfo  {journal} {Phys.
  Rev. C}\ }\textbf {\bibinfo {volume} {62}},\ \bibinfo {pages} {044307}
  (\bibinfo {year} {2000})}\BibitemShut {NoStop}%
\bibitem [{\citenamefont {Zhang}\ and\ \citenamefont
  {Niu}(2018)}]{PhysRevC.97.054302}%
  \BibitemOpen
  \bibfield  {author} {\bibinfo {author} {\bibfnamefont {W.}~\bibnamefont
  {Zhang}}\ and\ \bibinfo {author} {\bibfnamefont {Y.~F.}\ \bibnamefont
  {Niu}},\ }\href {\doibase 10.1103/PhysRevC.97.054302} {\bibfield  {journal}
  {\bibinfo  {journal} {Phys. Rev. C}\ }\textbf {\bibinfo {volume} {97}},\
  \bibinfo {pages} {054302} (\bibinfo {year} {2018})}\BibitemShut {NoStop}%
\bibitem [{\citenamefont {Zhang}\ and\ \citenamefont
  {Niu}(2017)}]{PhysRevC.96.054308}%
  \BibitemOpen
  \bibfield  {author} {\bibinfo {author} {\bibfnamefont {W.}~\bibnamefont
  {Zhang}}\ and\ \bibinfo {author} {\bibfnamefont {Y.~F.}\ \bibnamefont
  {Niu}},\ }\href {\doibase 10.1103/PhysRevC.96.054308} {\bibfield  {journal}
  {\bibinfo  {journal} {Phys. Rev. C}\ }\textbf {\bibinfo {volume} {96}},\
  \bibinfo {pages} {054308} (\bibinfo {year} {2017})}\BibitemShut {NoStop}%
\bibitem [{\citenamefont {Lisboa}\ \emph {et~al.}(2016)\citenamefont {Lisboa},
  \citenamefont {Malheiro},\ and\ \citenamefont
  {Carlson}}]{PhysRevC.93.024321}%
  \BibitemOpen
  \bibfield  {author} {\bibinfo {author} {\bibfnamefont {R.}~\bibnamefont
  {Lisboa}}, \bibinfo {author} {\bibfnamefont {M.}~\bibnamefont {Malheiro}}, \
  and\ \bibinfo {author} {\bibfnamefont {B.~V.}\ \bibnamefont {Carlson}},\
  }\href {\doibase 10.1103/PhysRevC.93.024321} {\bibfield  {journal} {\bibinfo
  {journal} {Phys. Rev. C}\ }\textbf {\bibinfo {volume} {93}},\ \bibinfo
  {pages} {024321} (\bibinfo {year} {2016})}\BibitemShut {NoStop}%
\bibitem [{\citenamefont {Ravli{\'{c}}}\ \emph {et~al.}(2023)\citenamefont
  {Ravli{\'{c}}}, \citenamefont {Y{\"u}ksel}, \citenamefont
  {Nik{\v{s}}i{\'{c}}},\ and\ \citenamefont {Paar}}]{Ravlic2023}%
  \BibitemOpen
  \bibfield  {author} {\bibinfo {author} {\bibfnamefont {A.}~\bibnamefont
  {Ravli{\'{c}}}}, \bibinfo {author} {\bibfnamefont {E.}~\bibnamefont
  {Y{\"u}ksel}}, \bibinfo {author} {\bibfnamefont {T.}~\bibnamefont
  {Nik{\v{s}}i{\'{c}}}}, \ and\ \bibinfo {author} {\bibfnamefont
  {N.}~\bibnamefont {Paar}},\ }\href {\doibase 10.1038/s41467-023-40613-2}
  {\bibfield  {journal} {\bibinfo  {journal} {Nature Communications}\ }\textbf
  {\bibinfo {volume} {14}},\ \bibinfo {pages} {4834} (\bibinfo {year}
  {2023})}\BibitemShut {NoStop}%
\bibitem [{\citenamefont {Y\"uksel}\ \emph {et~al.}(2019)\citenamefont
  {Y\"uksel}, \citenamefont {Marketin},\ and\ \citenamefont
  {Paar}}]{PhysRevC.99.034318}%
  \BibitemOpen
  \bibfield  {author} {\bibinfo {author} {\bibfnamefont {E.}~\bibnamefont
  {Y\"uksel}}, \bibinfo {author} {\bibfnamefont {T.}~\bibnamefont {Marketin}},
  \ and\ \bibinfo {author} {\bibfnamefont {N.}~\bibnamefont {Paar}},\ }\href
  {\doibase 10.1103/PhysRevC.99.034318} {\bibfield  {journal} {\bibinfo
  {journal} {Phys. Rev. C}\ }\textbf {\bibinfo {volume} {99}},\ \bibinfo
  {pages} {034318} (\bibinfo {year} {2019})}\BibitemShut {NoStop}%
\bibitem [{\citenamefont {Nikšić}\ \emph {et~al.}(2014)\citenamefont
  {Nikšić}, \citenamefont {Paar}, \citenamefont {Vretenar},\ and\
  \citenamefont {Ring}}]{NIKSIC20141808}%
  \BibitemOpen
  \bibfield  {author} {\bibinfo {author} {\bibfnamefont {T.}~\bibnamefont
  {Nikšić}}, \bibinfo {author} {\bibfnamefont {N.}~\bibnamefont {Paar}},
  \bibinfo {author} {\bibfnamefont {D.}~\bibnamefont {Vretenar}}, \ and\
  \bibinfo {author} {\bibfnamefont {P.}~\bibnamefont {Ring}},\ }\href {\doibase
  https://doi.org/10.1016/j.cpc.2014.02.027} {\bibfield  {journal} {\bibinfo
  {journal} {Computer Physics Communications}\ }\textbf {\bibinfo {volume}
  {185}},\ \bibinfo {pages} {1808} (\bibinfo {year} {2014})}\BibitemShut
  {NoStop}%
\bibitem [{\citenamefont {Ring}\ \emph {et~al.}(1984)\citenamefont {Ring},
  \citenamefont {Robledo}, \citenamefont {Egido},\ and\ \citenamefont
  {Faber}}]{RING1984261}%
  \BibitemOpen
  \bibfield  {author} {\bibinfo {author} {\bibfnamefont {P.}~\bibnamefont
  {Ring}}, \bibinfo {author} {\bibfnamefont {L.}~\bibnamefont {Robledo}},
  \bibinfo {author} {\bibfnamefont {J.}~\bibnamefont {Egido}}, \ and\ \bibinfo
  {author} {\bibfnamefont {M.}~\bibnamefont {Faber}},\ }\href {\doibase
  https://doi.org/10.1016/0375-9474(84)90393-2} {\bibfield  {journal} {\bibinfo
   {journal} {Nuclear Physics A}\ }\textbf {\bibinfo {volume} {419}},\ \bibinfo
  {pages} {261} (\bibinfo {year} {1984})}\BibitemShut {NoStop}%
\bibitem [{\citenamefont {Nik\ifmmode \check{s}\else
  \v{s}\fi{}i\ifmmode~\acute{c}\else \'{c}\fi{}}\ \emph
  {et~al.}(2002)\citenamefont {Nik\ifmmode \check{s}\else
  \v{s}\fi{}i\ifmmode~\acute{c}\else \'{c}\fi{}}, \citenamefont {Vretenar},
  \citenamefont {Finelli},\ and\ \citenamefont {Ring}}]{Niksic2002}%
  \BibitemOpen
  \bibfield  {author} {\bibinfo {author} {\bibfnamefont {T.}~\bibnamefont
  {Nik\ifmmode \check{s}\else \v{s}\fi{}i\ifmmode~\acute{c}\else \'{c}\fi{}}},
  \bibinfo {author} {\bibfnamefont {D.}~\bibnamefont {Vretenar}}, \bibinfo
  {author} {\bibfnamefont {P.}~\bibnamefont {Finelli}}, \ and\ \bibinfo
  {author} {\bibfnamefont {P.}~\bibnamefont {Ring}},\ }\href {\doibase
  10.1103/PhysRevC.66.024306} {\bibfield  {journal} {\bibinfo  {journal} {Phys.
  Rev. C}\ }\textbf {\bibinfo {volume} {66}},\ \bibinfo {pages} {024306}
  (\bibinfo {year} {2002})}\BibitemShut {NoStop}%
\bibitem [{\citenamefont {Tian}\ \emph {et~al.}(2009)\citenamefont {Tian},
  \citenamefont {Ma},\ and\ \citenamefont {Ring}}]{PhysRevC.80.024313}%
  \BibitemOpen
  \bibfield  {author} {\bibinfo {author} {\bibfnamefont {Y.}~\bibnamefont
  {Tian}}, \bibinfo {author} {\bibfnamefont {Z.-y.}\ \bibnamefont {Ma}}, \ and\
  \bibinfo {author} {\bibfnamefont {P.}~\bibnamefont {Ring}},\ }\href {\doibase
  10.1103/PhysRevC.80.024313} {\bibfield  {journal} {\bibinfo  {journal} {Phys.
  Rev. C}\ }\textbf {\bibinfo {volume} {80}},\ \bibinfo {pages} {024313}
  (\bibinfo {year} {2009})}\BibitemShut {NoStop}%
\bibitem [{\citenamefont {Egido}(1988)}]{PhysRevLett.61.767}%
  \BibitemOpen
  \bibfield  {author} {\bibinfo {author} {\bibfnamefont {J.~L.}\ \bibnamefont
  {Egido}},\ }\href {\doibase 10.1103/PhysRevLett.61.767} {\bibfield  {journal}
  {\bibinfo  {journal} {Phys. Rev. Lett.}\ }\textbf {\bibinfo {volume} {61}},\
  \bibinfo {pages} {767} (\bibinfo {year} {1988})}\BibitemShut {NoStop}%
\bibitem [{\citenamefont {Egido}\ and\ \citenamefont
  {Ring}(1993)}]{Egido_1993}%
  \BibitemOpen
  \bibfield  {author} {\bibinfo {author} {\bibfnamefont {J.~L.}\ \bibnamefont
  {Egido}}\ and\ \bibinfo {author} {\bibfnamefont {P.}~\bibnamefont {Ring}},\
  }\href {\doibase 10.1088/0954-3899/19/1/002} {\bibfield  {journal} {\bibinfo
  {journal} {Journal of Physics G: Nuclear and Particle Physics}\ }\textbf
  {\bibinfo {volume} {19}},\ \bibinfo {pages} {1} (\bibinfo {year}
  {1993})}\BibitemShut {NoStop}%
\bibitem [{\citenamefont {Goodman}(1984)}]{PhysRevC.29.1887}%
  \BibitemOpen
  \bibfield  {author} {\bibinfo {author} {\bibfnamefont {A.~L.}\ \bibnamefont
  {Goodman}},\ }\href {\doibase 10.1103/PhysRevC.29.1887} {\bibfield  {journal}
  {\bibinfo  {journal} {Phys. Rev. C}\ }\textbf {\bibinfo {volume} {29}},\
  \bibinfo {pages} {1887} (\bibinfo {year} {1984})}\BibitemShut {NoStop}%
\bibitem [{\citenamefont {Martin}\ \emph {et~al.}(2003)\citenamefont {Martin},
  \citenamefont {Egido},\ and\ \citenamefont {Robledo}}]{PhysRevC.68.034327}%
  \BibitemOpen
  \bibfield  {author} {\bibinfo {author} {\bibfnamefont {V.}~\bibnamefont
  {Martin}}, \bibinfo {author} {\bibfnamefont {J.~L.}\ \bibnamefont {Egido}}, \
  and\ \bibinfo {author} {\bibfnamefont {L.~M.}\ \bibnamefont {Robledo}},\
  }\href {\doibase 10.1103/PhysRevC.68.034327} {\bibfield  {journal} {\bibinfo
  {journal} {Phys. Rev. C}\ }\textbf {\bibinfo {volume} {68}},\ \bibinfo
  {pages} {034327} (\bibinfo {year} {2003})}\BibitemShut {NoStop}%
\bibitem [{\citenamefont {Dobaczewski}\ \emph {et~al.}(1996)\citenamefont
  {Dobaczewski}, \citenamefont {Nazarewicz}, \citenamefont {Werner},
  \citenamefont {Berger}, \citenamefont {Chinn},\ and\ \citenamefont
  {Decharg\'e}}]{PhysRevC.53.2809}%
  \BibitemOpen
  \bibfield  {author} {\bibinfo {author} {\bibfnamefont {J.}~\bibnamefont
  {Dobaczewski}}, \bibinfo {author} {\bibfnamefont {W.}~\bibnamefont
  {Nazarewicz}}, \bibinfo {author} {\bibfnamefont {T.~R.}\ \bibnamefont
  {Werner}}, \bibinfo {author} {\bibfnamefont {J.~F.}\ \bibnamefont {Berger}},
  \bibinfo {author} {\bibfnamefont {C.~R.}\ \bibnamefont {Chinn}}, \ and\
  \bibinfo {author} {\bibfnamefont {J.}~\bibnamefont {Decharg\'e}},\ }\href
  {\doibase 10.1103/PhysRevC.53.2809} {\bibfield  {journal} {\bibinfo
  {journal} {Phys. Rev. C}\ }\textbf {\bibinfo {volume} {53}},\ \bibinfo
  {pages} {2809} (\bibinfo {year} {1996})}\BibitemShut {NoStop}%
\bibitem [{\citenamefont {Besprosvany}\ and\ \citenamefont
  {Levit}(1989)}]{BESPROSVANY19891}%
  \BibitemOpen
  \bibfield  {author} {\bibinfo {author} {\bibfnamefont {J.}~\bibnamefont
  {Besprosvany}}\ and\ \bibinfo {author} {\bibfnamefont {S.}~\bibnamefont
  {Levit}},\ }\href {\doibase https://doi.org/10.1016/0370-2693(89)91504-9}
  {\bibfield  {journal} {\bibinfo  {journal} {Physics Letters B}\ }\textbf
  {\bibinfo {volume} {217}},\ \bibinfo {pages} {1} (\bibinfo {year}
  {1989})}\BibitemShut {NoStop}%
\bibitem [{\citenamefont {Egido}\ \emph {et~al.}(2000)\citenamefont {Egido},
  \citenamefont {Robledo},\ and\ \citenamefont {Martin}}]{PhysRevLett.85.26}%
  \BibitemOpen
  \bibfield  {author} {\bibinfo {author} {\bibfnamefont {J.~L.}\ \bibnamefont
  {Egido}}, \bibinfo {author} {\bibfnamefont {L.~M.}\ \bibnamefont {Robledo}},
  \ and\ \bibinfo {author} {\bibfnamefont {V.}~\bibnamefont {Martin}},\ }\href
  {\doibase 10.1103/PhysRevLett.85.26} {\bibfield  {journal} {\bibinfo
  {journal} {Phys. Rev. Lett.}\ }\textbf {\bibinfo {volume} {85}},\ \bibinfo
  {pages} {26} (\bibinfo {year} {2000})}\BibitemShut {NoStop}%
\bibitem [{\citenamefont {Brack}\ and\ \citenamefont
  {Quentin}(1974)}]{BRACK1974159}%
  \BibitemOpen
  \bibfield  {author} {\bibinfo {author} {\bibfnamefont {M.}~\bibnamefont
  {Brack}}\ and\ \bibinfo {author} {\bibfnamefont {P.}~\bibnamefont
  {Quentin}},\ }\href {\doibase https://doi.org/10.1016/0370-2693(74)90077-X}
  {\bibfield  {journal} {\bibinfo  {journal} {Physics Letters B}\ }\textbf
  {\bibinfo {volume} {52}},\ \bibinfo {pages} {159} (\bibinfo {year}
  {1974})}\BibitemShut {NoStop}%
\bibitem [{\citenamefont {Levit}\ and\ \citenamefont
  {Alhassid}(1984)}]{LEVIT1984439}%
  \BibitemOpen
  \bibfield  {author} {\bibinfo {author} {\bibfnamefont {S.}~\bibnamefont
  {Levit}}\ and\ \bibinfo {author} {\bibfnamefont {Y.}~\bibnamefont
  {Alhassid}},\ }\href {\doibase https://doi.org/10.1016/0375-9474(84)90421-4}
  {\bibfield  {journal} {\bibinfo  {journal} {Nuclear Physics A}\ }\textbf
  {\bibinfo {volume} {413}},\ \bibinfo {pages} {439} (\bibinfo {year}
  {1984})}\BibitemShut {NoStop}%
\bibitem [{\citenamefont {Chen}\ \emph {et~al.}(2005)\citenamefont {Chen},
  \citenamefont {Ko},\ and\ \citenamefont {Li}}]{PhysRevC.72.064309}%
  \BibitemOpen
  \bibfield  {author} {\bibinfo {author} {\bibfnamefont {L.-W.}\ \bibnamefont
  {Chen}}, \bibinfo {author} {\bibfnamefont {C.~M.}\ \bibnamefont {Ko}}, \ and\
  \bibinfo {author} {\bibfnamefont {B.-A.}\ \bibnamefont {Li}},\ }\href
  {\doibase 10.1103/PhysRevC.72.064309} {\bibfield  {journal} {\bibinfo
  {journal} {Phys. Rev. C}\ }\textbf {\bibinfo {volume} {72}},\ \bibinfo
  {pages} {064309} (\bibinfo {year} {2005})}\BibitemShut {NoStop}%
\bibitem [{\citenamefont {Centelles}\ \emph {et~al.}(2009)\citenamefont
  {Centelles}, \citenamefont {Roca-Maza}, \citenamefont {Vi\~nas},\ and\
  \citenamefont {Warda}}]{PhysRevLett.102.122502}%
  \BibitemOpen
  \bibfield  {author} {\bibinfo {author} {\bibfnamefont {M.}~\bibnamefont
  {Centelles}}, \bibinfo {author} {\bibfnamefont {X.}~\bibnamefont
  {Roca-Maza}}, \bibinfo {author} {\bibfnamefont {X.}~\bibnamefont {Vi\~nas}},
  \ and\ \bibinfo {author} {\bibfnamefont {M.}~\bibnamefont {Warda}},\ }\href
  {\doibase 10.1103/PhysRevLett.102.122502} {\bibfield  {journal} {\bibinfo
  {journal} {Phys. Rev. Lett.}\ }\textbf {\bibinfo {volume} {102}},\ \bibinfo
  {pages} {122502} (\bibinfo {year} {2009})}\BibitemShut {NoStop}%
\bibitem [{\citenamefont {Roca-Maza}\ and\ \citenamefont
  {Paar}(2018)}]{ROCAMAZA201896}%
  \BibitemOpen
  \bibfield  {author} {\bibinfo {author} {\bibfnamefont {X.}~\bibnamefont
  {Roca-Maza}}\ and\ \bibinfo {author} {\bibfnamefont {N.}~\bibnamefont
  {Paar}},\ }\href {\doibase https://doi.org/10.1016/j.ppnp.2018.04.001}
  {\bibfield  {journal} {\bibinfo  {journal} {Progress in Particle and Nuclear
  Physics}\ }\textbf {\bibinfo {volume} {101}},\ \bibinfo {pages} {96}
  (\bibinfo {year} {2018})}\BibitemShut {NoStop}%
\bibitem [{\citenamefont {De}\ and\ \citenamefont
  {Samaddar}(2012)}]{PhysRevC.85.024310}%
  \BibitemOpen
  \bibfield  {author} {\bibinfo {author} {\bibfnamefont {J.~N.}\ \bibnamefont
  {De}}\ and\ \bibinfo {author} {\bibfnamefont {S.~K.}\ \bibnamefont
  {Samaddar}},\ }\href {\doibase 10.1103/PhysRevC.85.024310} {\bibfield
  {journal} {\bibinfo  {journal} {Phys. Rev. C}\ }\textbf {\bibinfo {volume}
  {85}},\ \bibinfo {pages} {024310} (\bibinfo {year} {2012})}\BibitemShut
  {NoStop}%
\bibitem [{\citenamefont {Antonov}\ \emph {et~al.}(2017)\citenamefont
  {Antonov}, \citenamefont {Kadrev}, \citenamefont {Gaidarov}, \citenamefont
  {Sarriguren},\ and\ \citenamefont {de~Guerra}}]{PhysRevC.95.024314}%
  \BibitemOpen
  \bibfield  {author} {\bibinfo {author} {\bibfnamefont {A.~N.}\ \bibnamefont
  {Antonov}}, \bibinfo {author} {\bibfnamefont {D.~N.}\ \bibnamefont {Kadrev}},
  \bibinfo {author} {\bibfnamefont {M.~K.}\ \bibnamefont {Gaidarov}}, \bibinfo
  {author} {\bibfnamefont {P.}~\bibnamefont {Sarriguren}}, \ and\ \bibinfo
  {author} {\bibfnamefont {E.~M.}\ \bibnamefont {de~Guerra}},\ }\href {\doibase
  10.1103/PhysRevC.95.024314} {\bibfield  {journal} {\bibinfo  {journal} {Phys.
  Rev. C}\ }\textbf {\bibinfo {volume} {95}},\ \bibinfo {pages} {024314}
  (\bibinfo {year} {2017})}\BibitemShut {NoStop}%
\bibitem [{\citenamefont {Bethe}(1936)}]{PhysRev.50.332}%
  \BibitemOpen
  \bibfield  {author} {\bibinfo {author} {\bibfnamefont {H.~A.}\ \bibnamefont
  {Bethe}},\ }\href {\doibase 10.1103/PhysRev.50.332} {\bibfield  {journal}
  {\bibinfo  {journal} {Phys. Rev.}\ }\textbf {\bibinfo {volume} {50}},\
  \bibinfo {pages} {332} (\bibinfo {year} {1936})}\BibitemShut {NoStop}%
\end{thebibliography}%

\end{document}